# Learning Contextualized Semantics from Co-occurring Terms via a Siamese Architecture

**Ubai Sandouk**  UBAI.SANDOUK@MANCHESTER.AC.UK
*School of Computer Science*
*University of Manchester*
*Manchester, M13 9PL, UK*

**Ke Chen**[1]  CHEN@CS.MANCHESTER.AC.UK
*School of Computer Science*
*University of Manchester*
*Manchester, M13 9PL, UK*

**Abstract**

One of the biggest challenges in Multimedia information retrieval and understanding is to bridge the semantic gap by properly modeling concept semantics in context. The presence of out of vocabulary (OOV) concepts exacerbates this difficulty. To address the semantic gap issues, we formulate a problem on learning contextualized semantics from descriptive terms and propose a novel Siamese architecture to model the contextualized semantics from descriptive terms. By means of pattern aggregation and probabilistic topic models, our Siamese architecture captures contextualized semantics from the co-occurring descriptive terms via unsupervised learning, which leads to a concept embedding space of the terms in context. Furthermore, the co-occurring OOV concepts can be easily represented in the learnt concept embedding space. The main properties of the concept embedding space are demonstrated via visualization. Using various settings in semantic priming, we have carried out a thorough evaluation by comparing our approach to a number of state-of-the-art methods on six annotation corpora in different domains, i.e., MagTag5K, CAL500 and Million Song Dataset in the music domain as well as Corel5K, LabelMe and SUNDatabase in the image domain. Experimental results on semantic priming suggest that our approach outperforms those state-of-the-art methods considerably in various aspects.

**Keywords:** Contextualized Semantics, Descriptive terms, Siamese architecture, Out of Vocabulary, Semantic Priming

## 1 Introduction

*Multimedia information retrieval* (MMIR) is a collective terminology referring to a number of tasks involving indexing, comparison and retrieval of multimedia objects (Jaimes et al., 2005). As media content is created at an exponential rate, it has become increasingly difficult to manage even personal repositories of multimedia so as to make MMIR more and more demanding. Moreover, users expect certain levels of MMIR services from web service providers such as YouTube and Flickr. In addition, information processing tasks in fields such as medicine (Müller et al., 2004) and education (Chang, Eleftheriadis & Mcclintock, 1998) benefit enormously from advances in MMIR. In general, the most challenging problem in MMIR is the so-called *semantic*

---
[1] Corresponding author.

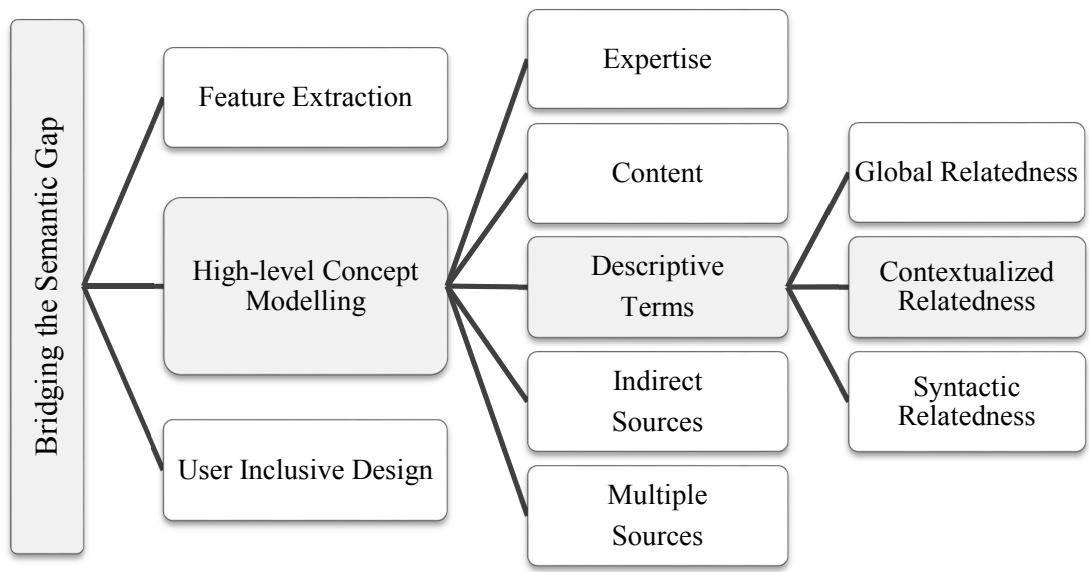

Figure 1: Main approaches and information sources used in bridging the semantic gap in MMIR.

*gap* (Smeulders et al., 2000), which stems from the difficulty in linking low-level media representation, e.g., computationally extractable features, to high-level semantic concepts describing the media content, e.g., human-like understanding. Bridging this gap has motivated a number of approaches including feature extraction (Lew et al., 2006), user-inclusive design (Schedl, Flexer & Urbano, 2013), and high-level context modeling (Marques, Barenholtz & Charvillat, 2011). By modeling concepts, the use of *semantics*, i.e., the representation of high-level concepts and their interactions, leads to improvements in MMIR applications as well as the interpretability of the retrieved results (Kaminskas & Ricci, 2012). As a result, semantics acquisition and representation are critical in bridging the semantic gap. The richness, meaningfulness and applicability of semantics rely primarily on the sources of concept-level relatedness information. Examples of such sources include manually constructed knowledge graphs or ontologies (Kim et al., 2008), automatically analyzed media content (Torralba, 2003) or well-explored collections of crowd-sourced descriptive terms or tags (Miotto & Lanckriet, 2012). Figure 1 summarizes main approaches to bridging the semantic gap and the commonly used sources of relatedness information.

In Figure 1, semantics obtained from *expertise* refers to manually enumerated intention(s) of individual concepts and their inter-relatedness. For instance, the meaning of a concept, such as activity 'upper arm bench', can be enumerated by a set of its attributes, such as 'arms back' and 'arms curl' (Sun et al., 2013). Expert-based semantics is interpretable and transferable but incurs extensive handcrafted work and the intrinsic difficulty in quantifying relatedness. Similarly, semantics obtained from the *content* refers to those generated by probing the media content for contextualizing information regarding the underlying concepts. For example, a relationship between the summary of a visual scene, e.g., seaside scene, and the objects within, e.g., boats and ships, can be attained to improve the object identification performance (Leong & Mihalcea, 2011; Torralba, Murphy & Freeman, 2010; Torralba, 2003). Unfortunately, content-based semantics may lack consistency and hence results in deterioration in the overall accuracy of the learnt semantics. In this circumstance, the contextualizing cues of the same concept appear in different forms and often mixed with other information components. For this reason, we argue that semantics learning should explore sources of relatedness information without involving the content itself. Similarly, semantics obtained from *indirect sources* refer to those semantics learnt from unobvious relatedness indicators, such as song co-occurrence in public playlists (Moore et al., 2012) or text/image adjacency online (Jing et al., 2014), where the incident of multimedia adjacency is assumed to imply relatedness. This is then used to infer concept-level relatedness. However, concept relatedness information is often entangled with irrelevant information components. Hence, it is more difficult to sift through a large amount of data in pursuit of concept relatedness information. Moreover, the contexts used in inferring indirect semantics are difficult to make learnt semantics interpretable and transferable. Apart from the aforementioned sources, semantics can also be obtained from *multiple sources*. For example, the introduction of relatedness-aware refinements with different sources of information over the concept detection task (Fink & Perona, 2003; Miotto & Lanckriet, 2012; Xiang et al., 2010) and the fusion of different high-level types of relatedness (Globerson et al., 2007; Mandel et al., 2011; Weston, Bengio & Hamel, 2011). Nevertheless, multiple sources of relatedness may not be complementary. Therefore, one needs to resolve incompatibilities among different sources with manual intervention and/or further investigation of additional data. Due to the complexity and unavailability of data in multiple sources, the learnt relatedness may be not sufficiently justifiable and difficult to use.

As an alternative form of information source, *descriptive terms,* including keywords, labels and other textual descriptions of media, have also been used in capturing the term-based semantics underlying co-occurring descriptive terms. Such semantics provides direct concept-level knowledge regarding the concerned multimedia objects. Typical applications include music crowd tagging services (Law, Settles & Mitchell, 2010) and multi-object image data set analysis (Rabinovich et al., 2007). Thanks to crowd-sourced annotation (Turnbull, Barrington & Lanckriet, 2008) and game-based tags collection (Law et al., 2007), large collections of descriptive terms are now accessible. Those term collections can be analyzed for occurring patterns to reveal concept-level relatedness and similarity. Term-based semantics is expected to be transferable since it is acquired from high-level concepts independent of any specific MMIR tasks. As depicted in Figure 1, semantics or relatedness of different types can be acquired from descriptive terms and will be reviewed in Section 2. It is worth stating that term-based semantics are different from linguistic semantics. First of all, descriptive terms are not only words but also symbols, abbreviations and complete sentences, e.g., "r'n'b" (musical style), "90s" (musical type), "stack of books" (visual concept), and so on. Next, descriptive terms may have a domain specific meaning different from their common linguistic meaning, e.g., "rock" is genre in music (not an earth substance) and "horn" is an instrument in music but is also a visual concept in images. Finally, the vocabulary used for descriptive terms is subject to change in time and cannot be fixed to represent a closed set of concepts. Those distinctions limit the usability of available linguistic resources such as linguistic dictionaries and generic word embedding from capturing term-based semantics. Although the information carried by descriptive terms is confined to the concepts expressed in textual form, we believe that the rich semantics conveyed in descriptive terms should be better explored and exploited to bridge the semantic gap.

By close investigation of various descriptive terms collections, we observe that terms could be used differently to represent various types of semantics and relatedness: a) a term may have multiple meanings and the intended meaning cannot be decided unless the term co-occurs with other coherent terms, e.g., the term "guitar" generally refers to an acoustic guitar when it co-occurs with terms like "strings", "classical", and so on, or to an electric guitar when it co-occurs with terms such as "metal", "rock", and so on; b) different terms may intend the exact same meaning regardless of context, e.g., "drums" and "drumset"; c) different terms may have either similar or different meaning depending on context, e.g., "trees" and "forest" convey similar concepts (many trees) and have similar meaning in context of natural scene (conveying a concept of many trees) but "tree" is by no means similar to "forest" when used in description of an urban scene; d) different terms may share partial meaning but have different connotations, e.g., "house" and "building" convey some similar concepts but "building" has a wider connotation; and e) co-occurring terms may not have their meanings in singularity or in pair but in group only, e.g., {"wing", "tail", "metallic"} together define a concept of an airplane while {"leg", "cat", "tail",…} collectively present a concept of a cat and its body parts. The observations described above indicate the complexity and the necessity of taking the context into account in semantic learning from terms. Obviously, simply counting co-occurrence (Rabinovich et al., 2007) is insufficient in modeling various types of semantics and relatedness in descriptive terms to capture accurate concepts, and more sophisticated techniques are required so that we can capture all the intended semantics or concepts and their relatedness in descriptive terms accurately.

In general, a set of $m$ terms, $\delta = \{\tau_i\}_{i=1}^m$, are often used collectively to describe the semantics underlying a single multimedia object where $\tau_i$ is a descriptive term and $\delta$ is the collective notation of the $m$ terms, named *document* hereinafter. Furthermore, all $m$ terms appearing in a document $\delta$ are dubbed as *accompany* terms. Our observation reveals that for a specific term $\tau_i$ in a document $\delta$, the accompany terms jointly create its contextual niche, named *local context,* that helps inferring the accurate intended meaning of $\tau_i$ in that situation. In other words, the term along its local context uniquely define a concept of the accurate meaning. By taking such local contexts into account, we would learn a new type of relatedness between terms, named *contextualized relatedness*, by exploring terms' co-occurrence in different documents in a collection. Unlike the *global relatedness* where relatedness of terms is fixed irrespective of their local contexts, the contextualized relatedness of two terms is subject to change in the presence of different local contexts. In order to represent such contextualized semantics, we would embed all terms in a concept representation space that reflects the contextualized relatedness of terms. Formally, this emerging problem is formulated as follows: given a term $\tau$ and its accompany terms in $\delta$, we would establish a mapping: $(t(\tau), l(\tau|\delta)) \to CE(\tau|\delta)$, where $t(\tau)$ and $l(\tau|\delta)$ are the feature vectors of the term $\tau$ and its local context in $\delta$ and $CE(\tau|\delta)$ is a concept embedding representation of $\tau$ given its local context in $\delta$, so that the contextualized semantic similarity of terms be properly reflected via a distance metric in the concept embedding representation space. This is a challenging problem due to the actual facts as follows: a) terms get their meaning in groups rather than in singularity or in pair; b) it is unclear how to capture intrinsic context in terms; and c) terms that are not seen in training may appear in application runtime and hence may confuse a semantic learning model, this issue is known as *out-of-vocabulary* (OOV) issue in literature. Nevertheless, solving this problem brings us closer to bridging the semantic gap as a solution to this problem not only provides a term-level contextualized semantic representation, named *concept embedding* (CE) hereinafter, for a term to grasp an accurate concept as well as contextualized concept relatedness but also the representations of co-occurring terms in a document collectively form a novel document-level representation precisely modeling the concepts in groups as well as subtle differences among those coherent concepts. Furthermore, the **CE** representation learnt from descriptive terms would facilitate a number of non-trivial applications including different MMIR tasks, e.g., auto-annotation of multimedia objects by mapping from the low-level visual/acoustic features onto the CE space, semantic retrieval by using the embedding representations as indexing terms, generating useful recommendations on both term and document levels in a recommendation system, and zero-shot learning in different multimedia classification tasks, e.g., object recognition and music genre classification.

In order to tackle the problem described above, we propose a novel Siamese architecture (Bromley et al., 1993) and a two-stage learning algorithm to capture contextualized semantics from descriptive terms. The proposed Siamese architecture learns the contextualized semantic embedding in an unsupervised way. The resultant **CE** representation space embeds different descriptive terms so that their contextualized semantic relatedness is reflected by their Euclidean distances. In this **CE** representation space, one term thus tends to co-locate with all the accompany terms appearing in its local context or co-occurring terms in the same document. As a result, our approach leads to multiple representations for a single term that appears in various documents. The semantics learnt this way are naturally generic yet transferable as they do not rely on any specific MMIR tasks. Depending on the nature of descriptive terms used in practice, the semantics acquired from some training collections may also be domain specific. With different training and test corpora, we would verify the above-mentioned transferability and domain-specific properties of the **CE** representations generated by the Siamese architecture in our experiments of various settings.

Our main contributions in this paper are summarized as follows: a) we formulate a problem for learning contextualized semantics from co-occurring descriptive terms and propose a novel Siamese architecture and a two-stage learning algorithm to be a solution for this problem; b) we propose two treatments based on our **CE** representation to address the CE issues regarding OOV terms; c) we demonstrate the main properties of the **CE** representations via visualization; and d) by means of semantic priming, a well-known information retrieval task, we thoroughly evaluate the performance of the **CE** representations generated by our Siamese architecture with a number of various settings and compare it to those generated by several state-of-the-art semantic learning methods.

The rest of the paper is organized as follows. Section 2 reviews the related work in terms of learning different relatedness from descriptive terms. Section 3 describes feature extraction of term and local context required in our contextualized semantic learning. Section 4 presents our Siamese architecture and learning algorithms. Section 5 describes the experiments on the CE learning with our Siamese architecture and Section 6 presents experimental settings and results in semantic priming. Section 7 discusses relevant issues and the last section draws conclusions.

## 2   Related Work

In this section, we review relevant works in learning semantics from descriptive terms regardless of any specific multimedia tasks. In terms of semantics learnt from descriptive terms, those approaches fall into one of three different categories: global relatedness, syntactic relatedness and contextualized relatedness as shown in Figure 1.

### 2.1   Global Relatedness

Global relatedness refers to the relatedness between pairs of terms that does not take any context into account. Below we review two main methodologies of learning global relatedness from descriptive terms: statistics-based and graph-based methodologies.

Aggregation (Markines et al., 2009) is a statistical-based method that focuses on pairwise co-occurrence of terms in the training data set and is sometimes named co-occurrence analysis. By considering all training documents, aggregation works on a document-term matrix $\Upsilon$ where the presence or absence of each term in each document is represented as binary or frequency indicator (Singhal, 2001). Thus, a column of document-term matrix $\Upsilon$ forms a feature vector of one term. As relatedness between any pair of terms is most likely reflected in their pair-wise pattern of use, this relatedness can be estimated by measuring the distance between their corresponding feature vectors. Hence, the relatedness between any pair of terms can be learnt from a training set with statistical measures. As the relatedness is obtained from the co-occurrence information across an entire data set, they are not affected by the local context in a document. As an extension, the term-to-term relatedness matrix achieved with all the pair-wise relatedness may be analyzed further with the Principle Component Analysis (PCA), which results in a representation for each descriptive term in a lower dimensional space via removing unwanted co-occurrence redundancy and noise. According to (Lebret et al., 2013), this extension yields improved performance in a specific linguistic task: movie review sentiment evaluation. Nevertheless, such extension seems quite sensitive to preprocessing and tunable parameters. Similarly, Mandel et al. (2011) proposed an *info*rmation *theo*retic inspired (InfoTheo) method that yields a smoothed document representation in the document-term $\Upsilon$ matrix. InfoTheo directly alters the values in favor of terms that appear together frequently across an entire data set. This smoothed representation is later aggregated in order to generate a term-to-term relatedness matrix. Nevertheless, this smoothing process introduces more parameters and hence results in a heavier burden in parameter tuning.

Another statistic-based method is Latent Semantic Indexing (LSI) (Deerwester et al., 1990), a popular technique in text information retrieval. LSI is used to analyze collections of documents with large vocabulary or descriptive terms in our terminology. In LSI, the document-term matrix $\Upsilon$ is decomposed using Singular Value Decomposition (SVD), or two-mode factor analysis, as $\Upsilon^T = U\Sigma V^T$. The two orthogonal matrices $U$ and $V$ correspond to the two subspaces to be modeled: the terms subspace and the document subspace, respectively. The dimensionality of the term subspace is controlled by limiting the entries of the diagonal matrix $\Sigma$, i.e., keeping the first few diagonal entries and setting the rest to zero. This decomposition generates an approximation of $\Upsilon$ with the smallest reconstruction error. It also has the property of uncovering terms' usage patterns collectively. The rows of the resultant $U$ matrix can readily be used as the corresponding representations of the descriptive terms. The relatedness between the terms is measured by the cosine similarity between their corresponding vectors (Levy & Sandler, 2008). Unfortunately, LSI generally suffers from poor generalizability to new terms as well as to new documents in application.

Graph-based models rely on a graph representation where terms are mapped on nodes and edges between nodes indicate pair-wise relatedness. Establishing such a graph representation can be done either manually or by some relatedness analysis (Hueting, Monszpart & Mellado, 2014; Kim et al., 2008; Wang et al., 2010). Apart from those handcrafted, the graph may be constructed via statistical analysis similar to those methods described above. However, graph-based models are subject to capacity limitation; any additional node representing a new term has to be introduced manually in graph revision. Moreover, an edge representing a level of relatedness usually has a fixed cost that does not take the local context into account. Therefore, graph-based models are often thought of as mere handcrafted dictionaries of semantics.

In summary, those approaches to learning global relatedness do not address the issue of the contextualized semantics studied in this paper. However, those approaches yield a term-level semantic representation and often work very efficiently. In our work, we would explore such approaches in generating a semantic representation of terms required by our learning model (see the next section for details).

## 2.2 Syntactic Relatedness

In natural languages, context is explicitly present in the order of the words, i.e., syntactic context. This dependency between sequences of words helps capture the words' relatedness in context and consequently the understanding of basic linguistic meanings.

For capturing the syntactic relatedness, distributed language models (Mikolov et al., 2010; Collobert et al., 2011; Mikolov et al., 2013) were recently proposed. Such models learn syntactic relatedness from linguistic corpora and yield a distributed semantic space where all words are embedded properly based on their syntactic similarity (Mikolov et al., 2010). During learning, an architecture is trained to predict a missing word given some context, i.e., nearby words, or to predict possible context words given one word. If trained properly, words that can be used interchangeable without breaking language rules, i.e., syntactically close words, will have close embedding vectors. Those models have been attracting increasing attention due to their simplicity and capacity in providing generic semantics for various application tasks (Frome et al., 2013; Mikolov et al., 2013). Moreover, Pennington and Manning (2014) showed how to combine the advantages of simple PCA models with this syntactic relatedness by careful analysis of the ratios of co-occurrence probabilities between pairs of words appearing in each other's neighborhood. Relaxing the syntactic dependency assumption in such models reverts to co-occurrence analysis with more elaborate context constraints.

Although language models yield a contextualized representation, they entirely rely on the syntactic context and hence do not seem applicable to descriptive terms in a document where there is no synthetic dependency and the co-occurring terms may describe a multimedia object regardless of their orders. Nevertheless, such techniques can be employed as a baseline in a thorough evaluation of the contextualized semantics learned from descriptive terms studied in this paper.

## 2.3 Contextualized Relatedness

Motivated by syntactic relatedness, pairs of terms are permitted to exhibit varying relatedness levels depending on their contexts. Works in this stream focus on document-level representation where patterns of terms' use are captured in a document-level representation. Consequently, measuring similarity between documents may be straightforward while terms' meaning and their relatedness are difficult to estimate. This often hinders the applicability of such learnt models as a generic semantics provider. Here, we review a number of approaches that can potentially capture the contextual relatedness studied in this paper but end up with a document-level representation.

Topic models are a class of statistical methods used for semantics analysis and modeling. In general, a topic model makes use of latent processes to capture occurrence patterns in the form of statistical distributions over observed terms, often called topics. When a specific term appears in more than one document, which exhibits different patterns of joint use with other terms, it might be associated with more than one topic and hence suggests its different meanings stochastically. Those multiple term-topic associations eventually allow for different levels of relatedness between terms to emerge. Latent Direchlet Allocation (LDA) (Blei, Ng & Jordan, 2003) and Probabilistic Latent Semantic Analysis (PLSA) (Hofmann, 1999) are the two most prominent topic models used in text and natural language processing.

In LDA and PLSA, a set of independent topics $\Phi$ are used to softly cluster the documents based on the used terms. During learning, the process estimates scalar priors B for the Dirichlet distribution in LDA or Multinomial distribution in PLSA that models the topics as distributions over terms as well as the scalar prior $B^0$ that models the topics distribution itself. After training, the posterior probability of all the topics given a tag and the topic probability given a document can be estimated with the trained models. Given a term $\tau$, the posterior probability of a topic $\phi_c \in \Phi$ is $p(\phi_c|\tau) \sim p(\tau|\phi_c)p(\phi_c)$, where $p(\tau|\phi_c) \sim Categorical(Dirichlet(B_c))$ in LDA or $p(\tau|\phi_c) \sim Categorical(B_c)$ in P-LSA and $p(\phi_c) \sim Dirichlet(B^0)$ in LDA or $p(\phi_c) \sim Uniform(B^0)$ in P-LSA. Given a document $\delta$, the topic probability is $p(\phi_c|\delta) \sim p(\phi_c) \prod_{\tau \in \delta} p(\tau|\phi_c)$.

While LDA or PLSA capture some contextualized semantics, such models merely provide a summary of a document in form of a mixture of topics; they capture ad hoc relatedness but do not provide term-to-term relatedness explicitly. As a result, the relatedness between a pair of terms $\tau_1$ and $\tau_2$ has to be estimated under a specific topic distribution: $\theta(\delta) = \{p(\theta_c) = p(\phi_c|\delta)\}_{c=1}^{|\Phi|}$. Assuming equal priors for all the terms, the relatedness between two terms, $\tau_1$ and $\tau_2$, may be measured by the Kullback–Leibler (*KL*) divergence:

$$KL(\tau_1, \tau_2|\theta(\delta)) = \sum_{c=1}^{|\Phi|} (p(\theta_c|\tau_1) - p(\theta_c|\tau_2)) \cdot \left(\log\left(\frac{p(\theta_c|\tau_1)}{p(\theta_c|\tau_2)}\right)\right)$$

$$= \sum_{c=1}^{|\Phi|} p(\theta_c) \left(\frac{p(\tau_1|\theta_c)}{p(\tau_1)} - \frac{p(\tau_2|\theta_c)}{p(\tau_2)}\right) \cdot \left(\log\left(\frac{p(\tau_2).p(\tau_1|\theta_c)}{p(\tau_1).p(\tau_2|\theta_c)}\right)\right)$$

$$= \sum_{c=1}^{|\Phi|} \frac{p(\theta_c)}{p(\tau_1)} (p(\tau_1|\theta_c) - p(\tau_2|\theta_c)) \cdot \left(\log\left(\frac{p(\tau_1|\theta_c)}{p(\tau_2|\theta_c)}\right)\right). \quad (1)$$

Thus, the term-topic relatedness learnt by LDA or PLSA implicitly contextualizes the relatedness between terms under different topic distributions. Without any consideration of term-to-term relatedness directly, however, ad hoc semantics yielded by LDA or PLSA is simply encoded in a collective term representation rather than a concept-level relatedness representation required by any solution to our problem described in Section 1. The effectiveness of topic models in learning semantics from descriptive terms has been evaluated in (Law, Settles & Mitchell, 2010) and (Levy & Sandler, 2008). They reported increased accuracy over global relatedness models when performing the auto tagging task directly from music. In their settings, this task does not require a detailed term-to-term relatedness measure and hence a topic-to-term relatedness offered by topic models is sufficient in this auto-tagging task.

Another method for introducing a context is using Conditional Restricted Boltzmann Machines (CRBM) in (Mandel et al., 2011). CRBM (Taylor, Hinton & Roweis, 2007) is a variant of the traditional RBM (Hinton, 2002), a generative model that has two layers of probabilistic units: visible layer $\boldsymbol{v}$ and hidden layer $\boldsymbol{h}$. The units in visible and hidden layers are fully connected via a weight matrix $U$, and vectors $\boldsymbol{d}$ and $\boldsymbol{c}$ are the biases in visible and hidden layers, respectively. Given that RBM is an energy based model, the energy function is defined by

$$E(\boldsymbol{v}, \boldsymbol{h}) = -\boldsymbol{h}^T U \boldsymbol{v} - \boldsymbol{d}^T \boldsymbol{v} - \boldsymbol{c}^T \boldsymbol{h}.$$

The model is trained by increasing the probability of patterns seen in training. In other words, minimizing the system's free energy given by

$$F(\boldsymbol{v}) = -\log \sum_{\boldsymbol{h}} e^{-E(\boldsymbol{v},\boldsymbol{h})}.$$

The optimization can be accomplished by the contrastive divergence (CD) algorithm (Hinton, 2002) that implicitly samples the hidden distribution to estimate the free energy.

In CRBM (Mandel et al., 2011), the energy function used in RBM is modified to take the condition $\boldsymbol{a}$ into account with an additional visible layer connected to the original visible layer via a weight matrix $W$. As a result, the energy function in CRBM is defined by

$$E(\boldsymbol{v}, \boldsymbol{h}, \boldsymbol{a}) = -\boldsymbol{h}^T U \boldsymbol{v} - \boldsymbol{v}^T W \boldsymbol{a} - \boldsymbol{d}^T \boldsymbol{v} - \boldsymbol{c}^T \boldsymbol{h}.$$

CRBM can also be trained by the CD algorithm. When it is used in learning semantics from descriptive terms, the observed vector $\boldsymbol{v}$ is set to be the *Bag of Words* (BoW) binary representation (Harris, 1954) of a considered document. One binomial unit is used per vocabulary term. The hidden vector $\boldsymbol{h}$ is set to be a vector of binomially distributed variables that captures global occurrence patterns. The condition $\boldsymbol{a}$ is set to be a one-hot representation of the training documents where exactly one unit is used as the index representing the considered document out of all the documents. The collection of terms used by other users for the same document is also used as a different condition in (Mandel et al., 2011). Such conditions depend on data availability. During test, the term-to-term relatedness is measured by co-activation between a query term and all other available terms. In this process, the unit corresponding to one query term is clamped "on" (as well as setting the relevant condition units) and sampling chains for a large number of times are undertaken. Eventually, the average co-activation level of each visible unit encodes relatedness of its corresponding term to the query term under the context conditions. While CRBM provides a smoothed relevance of each term to the considered document, the captured semantics is limited to a document-level representation rather than concept-level relatedness studied in this paper. In particular, this approach suffers from a fundamental weakness as it does not lead to a deterministic continuous semantic embedding representation required by miscellaneous real applications.

Association rules (Agrawal, Imieliński & Swami, 1993) of the form {"computer", "apple"}→ {"mac"}, meaning that the presence of terms "computer" and "apple" implies the presence of term "mac", can be "mined" or learnt from the training data to represent relatedness between complete sets of terms. Patterns of co-occurrence are captured in interpretable dependency rules. Yang et al. (2010) suggested the use of mining association rules from image tagged data for the tag completion task. Unfortunately, the term-to-term relatedness is neither considered nor generated so that only group level rules are expressed. A lack of the term-to-term relatedness significantly limits the applicability of those mined rules. Moreover, the contextualizing information for a specific set of terms may differ in different rules. This inconsistency causes difficulty in understanding contexts and generalizing the mined rules to new data sets.

In summary, the existing work for learning semantics from descriptive terms focuses on relatedness of specific types and does not sufficiently address the issues appearing in our formulated problem. To verify our argument, we have used all the approaches reviewed above apart from the association rules as baselines in our experiments (see Section 6 for details).

# 3 Term, Local Context and Document Representation

In this section, we describe feature extraction for the descriptive term, local context and document representations employed in our approach to facilitate the presentation of our proposed Siamese architecture in the next section.

## 3.1 Term Representation

In general, terms can be characterized by either ID-based or statistics-based representations. An ID-based representation uses a scheme directly linked to the term's ID in symbol form, i.e., a separate entity for each term. A statistics-based representation uses statistical analysis of the use of the tag within the data set. The ID-based term representations are often used in previous contextualized models, e.g., the CRBM model (Mandel et al., 2011). However, the capacity of an ID-based representation may be limited so that it is difficult to accommodate new terms, and it may also impose an unnatural order on the terms, e.g., a numerical ID index representation. Therefore, we employ a statistics-based representation (Markines et al., 2009) where each descriptive term is represented as a summary of its pair-wise use with all terms over an entire training data set. This summary encodes the global relatedness among pairs of terms which will later be augmented by the local context to form a raw concept representation of a term in context.

To achieve the statistics-based representation, we start from the training document-term matrix with binary entries described in Section 2.1. Those entries represent the use of terms in the corresponding documents. In our work, we do not eliminate any terms in a training data set as we believe that the entire collection of terms in different documents forms a coherent meaning niche conveying proper local contexts collectively. The document-term binary matrix is re-weighted using $tfidf$ which highlights those rarely used or highly descriptive terms. Given a vocabulary of descriptive terms $\Gamma$ and a training data set $\Delta$, the binary *term frequency* of the presence of the term $\tau \in \Gamma$ in a document $\delta \in \Delta$ is found in the corresponding entry in the document-term matrix:

$$tf(\tau, \delta) = \begin{cases} 1 & when\ \tau\ appears\ in\ \delta \\ 0 & otherwise \end{cases}.$$

The rarity of a term $\tau$ in the collection is achieved by the inverted document frequency $idf(\tau)$:

$$idf(\tau) = log\left(\frac{|\Delta|}{1 + |\{\delta : \tau\ appears\ in\ \delta\}|}\right),$$

where $|.|$ is the cardinality of a set. Then the $tfidf$ weight is defined as:

$$tfidf(\tau, \delta) = tf(\tau, \delta) \times idf(\tau).$$

After reweighting the matrix, each term is described using all $tfidf$ values of its use and is represented by its usage vector $\boldsymbol{u}(\tau) = \{tfidf(\tau, \delta_i)\}_{i=1}^{|\Delta|}$. Then, the global relatedness between two terms $\tau_1$ and $\tau_2$ is obtained by aggregation with the dot product:

$$T(\tau_1, \tau_2) = <\boldsymbol{u}(\tau_1), \boldsymbol{u}(\tau_2)>$$

Thus, a term is represented by a feature vector of $|\Gamma|$ features consisting of its global relatedness to all terms in the training data set:

$$\boldsymbol{t}(\tau) = \{T(\tau, \tau_i)\}_{i=1}^{|\Gamma|}. \tag{2}$$

## 3.2 Local Context Representation

As described in Section 1, the local context of each term is acquired by considering all the co-occurring terms in the same document. In general, each document consists of a coherent set of descriptive terms. Without human knowledge and intervention, it is impossible to split accompany terms in a document into coherent subgroups of terms. Therefore, a local context representation has to be obtained from the entire document of co-occurring terms. An ideal local context representation should be semantically consistent across documents and easy to estimate in real applications, e.g., auto-tag annotation. Obviously, for a document, its one-hot representation out of the document list is unable to measure similarity between documents and is also subject to the generalization limitation. In recent work of Law, Settles & Mitchell (2010), Latent Direchlet Allocation (LDA) was used to represent terms in form of topics collectively and then a model was trained to map the acoustic content onto the topical representation to facilitate MMIR. Motivated by their work, we employ LDA to represent the local context in our work due to the generality of LDA in representing patterns of collective use. It is worth mentioning that there are alternative models for local context representations, e.g., semantic hierarchies, PLSA or any other topic model. Here, we emphasize that a local context representation used in our work is not equivalent to the complete document itself but a semantically coherent summary of the document.

To achieve the local context representation with LDA (see Section 2.3 for more details), a set of topics $\Phi$ is assumed to softly cluster the documents based on the used terms within each document. During training, the process estimates scalar priors B for the Dirichlet distributions to model the topics as distributions over terms as well as the scalar prior $B^0$ to model the topic distribution itself. Once the training is completed, the probability of observing a term $\tau \in \Gamma$ given a specific topic $\phi \in \Phi$ follows

$$p(\tau|\phi) \sim Categorical(Dirichlet(B)), \text{ where } p(\phi) \sim Dirichlet(B^0).$$

This means that the probability of one term identifying a specific topic follows $p(\phi|\tau) \sim p(\tau|\phi)p(\phi)$ and the likelihood of a topic given a complete document $\delta$ is:

$$p(\phi|\delta) \sim p(\phi) \prod_{\tau \in \delta} p(\tau|\phi).$$

Given the term and accompany terms, the local context of a term is thus represented by a feature vector of $|\Phi|$ features corresponding to $|\Phi|$ topic distribution output:

$$\boldsymbol{l}(\tau|\delta) = \{l_c(\delta)\}_{c=1}^{|\Phi|}, \quad l_c(\delta) = p(\phi_c|\delta). \tag{3}$$

## 3.3 Document Representation

Apart from term and local context representations, a representation of an entire document is also required in our approach presented in next Section. In our work, we adopt the Bag of Words (BoW) representation of a document $\delta$, denoted by $\boldsymbol{BoW}(\delta)$, a binary sparse feature vector of $|\Gamma|$ entries for a given vocabulary $\Gamma$ where entry $i$ corresponds to a specific term $\tau_i$. Thus:

$$\boldsymbol{BoW}(\delta)[i] = \begin{cases} 1 & \text{when } \tau_i \text{ appears in } \delta \\ 0 & \text{otherwise} \end{cases}. \tag{4}$$

In summary, we employ the $tfidf$-based aggregation as our term representation to encode the global term-to-term relatedness, the LDA as our local context representation to summarize semantic coherence in different documents and the BoW as the document representation in our learning model presented in the next Section.

# 4 Model Description

In this Section, we come up with a solution to the problem described in Section 1. We first describe our motivation behind our proposed Siamese architecture and then present its architecture and a two-stage algorithm for learning contextualized semantics from descriptive terms. Finally, we propose two methods to deal with the OOV terms contextualized semantic embedding based on the representation space generated by our Siamese architecture.

## 4.1 Motivation

As described in Section 1, we aim to tackle an issue that has not been fully addressed by the previous work in learning semantics from descriptive terms. In the previous work reviewed in Section 2, either the term-to-term relatedness is captured without taking the local context into account or the context is modeled on a document level. Unlike previous work, we encounter a challenge where a term and its local context are simultaneously taken into account to learn the contextualized term-to-term relatedness embedded in a semantic representational space. By looking into the nature of this problem, we would like to come up with a solution by carrying out two subsequent tasks.

In general, an ideal representation of semantics allows similar concepts to associate with one another seamlessly; a concept should be easily inferred from its related or coherent concepts. Motivated by the argument that learning a simple yet relevant auxiliary task could facilitate learning semantic embedding (Bottou, 2014), we can comply with this requirement by fulfilling a simple yet generic task of *predicting all the accompany terms in a document from the representations of a constitutional term and its local context* described in Section 3. If a learning model of latent variables is employed to carry out this task, we expect that the latent variables form a representational space that encodes the semantic information of coherent terms at a concept level. As such a representation also needs to resolve the highly nonlinear relationship between terms and their contexts to predict accompany terms, a deep neural network of hidden layers would be a powerful candidate for this task.

While the representation generated by carrying out that prediction task encodes the semantic information conveyed in coherent terms, it may not provide the proper term-to-term relatedness in context. Therefore, we need to perform a further task based on the initial semantic representation achieved in the prediction task; i.e., *learning a proper distance metric for the pairwise contextualized relatedness of concepts* to enhance the semantic representational space. To carry out this task, we would develop a variant of Siamese architecture (Bromley et al., 1993) consisting of two identical deep neural networks used in the earlier prediction task to be its subnetworks. By taking all possible concept relations between a pair of terms along with their local contexts into account during learning, we expect that all concepts reflected by terms in the presence of local contexts are embedded properly so that a pair of coherent concepts with the same context can be co-located in the representational space with minimal distance and other concepts can be distant in reflection of their contextualized relatedness.

By accomplishing two learning tasks, we anticipate that all contextualized concepts are embedded in a distributed representation space. This idea motivates us to develop a Siamese architecture and a two-stage unsupervised learning algorithm to tackle the problem of learning contextualized semantics from descriptive terms. To carry out this idea, we have to also address a number of non-trivial technical issues as presented in the rest of this section.

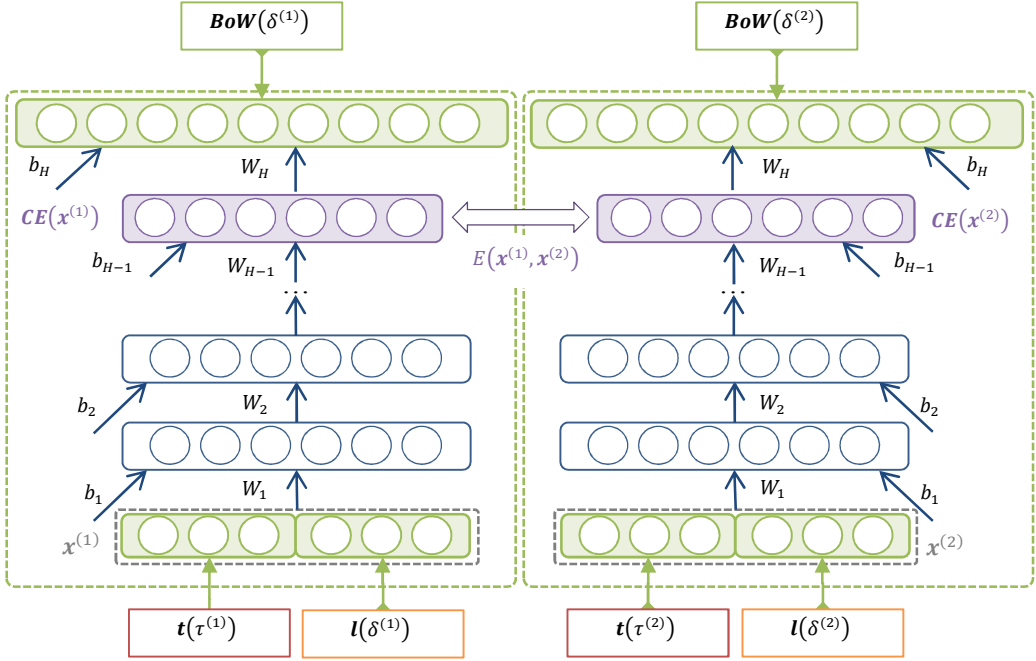

Figure 2: The proposed Siamese architecture for learning contextualized semantics from descriptive terms.

## 4.2 Architecture

As illustrated in Figure 2, the proposed Siamese architecture consists of two identical subnetworks. Each subnetwork is a feed-forward neural network composed of $H-1$ hidden layers and two visible layers marked in green; i.e., input and output layers. Each subnetwork receives the representations of a term $t(\tau)$ and its local context $l(\tau|\delta)$, collectively denoted by $x(\tau,\delta) = (t(\tau), l(\tau|\delta))$, as input and outputs a prediction of the BoW representation of all the terms in $\delta$, $BoW(\delta)$, in the document $\delta$ from which $t(\tau)$ and $l(\tau|\delta)$ were extracted. Two subnetworks are coupled to work together and trained via a two-stage learning procedure.

In the first stage, one subnetwork is trained to carry out the prediction task for an initial semantic embedding. As a result, this subnetwork is trained to predict the BoW representation of a document, $BoW(\delta)$, from the input features of a tag $\tau$ and its local context in $\delta$, $x(\tau,\delta)$. After the first-stage learning, the output of the $(H-1)^{th}$ hidden layer is used as an initial contextualized semantic representation for concepts conveyed by terms and their local contexts. We refer to this representation as *concept embedding* ($CE$) throughout the paper.

In the second stage, we couple two identical trained subnetworks and train two subnetworks simultaneously to revise the initial semantic embedding towards embedding the proper contextualized term-to-term or concept-to-concept relatedness. The learning in this stage is done via *distance learning* working on further constraints required by a proper distance metric in contextualized semantic embedding. During the distance learning, two subnetworks work together to deal with different situations regarding all possible types of input to two subnetworks. For regularization, each subnetwork is also trained simultaneously in this stage to perform the prediction task in order to avoid unnecessary changes to initial semantic representation achieved in the first stage with a multi-objective optimization process.

After the two-stage learning is completed, we achieve two identical subnetworks. Then one of them is used in mapping a term plus its local context to the CE space to form its contextualized semantic representation.

### 4.3 Learning Algorithm

To facilitate the presentation of our learning algorithm, we first describe our notation system. For layer number $h$ in a subnetwork, the output is

$$\boldsymbol{z}_h(\boldsymbol{x}) = \boldsymbol{f}(W_h \cdot \boldsymbol{z}_{h-1}(\boldsymbol{x}) + \boldsymbol{b}_h), 1 \leq h \leq H,$$

where $W_h, \boldsymbol{b}_h$ are the weights and bias vectors for the $h^{th}$ layer of the network, $\boldsymbol{f}(.)$ is the element-wise hyperbolic tangent function:

$$\boldsymbol{f}(x) = \frac{e^x - e^{-x}}{e^x + e^{-x}}.$$

We stipulate that $\boldsymbol{z}_0(\boldsymbol{x}) = \boldsymbol{x}$ indicates the input layer, $\boldsymbol{CE}(\boldsymbol{x}) = \boldsymbol{z}_{H-1}(\boldsymbol{x})$ is the contextualized semantic representation vector, i.e., the output of the $(H-1)^{th}$ hidden layer, and $\hat{\boldsymbol{y}}(\boldsymbol{x}) = \boldsymbol{z}_H(\boldsymbol{x})$ is the prediction vector yielded by the output layer. Hereinafter, we shall drop all the explicit parameters to simplify the presentation, e.g., $\boldsymbol{y}_k$ is an abbreviated version of $\boldsymbol{y}_k(\boldsymbol{x}_k(\tau, \delta))$ and $y_k[j]$ denotes the $j^{th}$ entry of the vector $\boldsymbol{y}_k$.

#### 4.3.1 Training Data

For learning, we need to create training examples based on different documents in a term collection or collections used for training. Given a training document $\delta$ consisting of $m$ co-occurring terms, we create $m$ training examples where each example is a focused term in document with the same local context, i.e., the $m$ terms in the document $\delta$. The prediction target for all the $m$ examples is the same, i.e. the document representation of this training document $\boldsymbol{BoW}(\delta)$. We observed that in training for the prediction, the local context may predominate the initial semantic embedding and hence cause all terms in the same document to have very similar representations regardless of whether they are meaningfully coherent. As a result, we have to tackle this issue by introducing *negative examples*. Given a training document $\delta$, we synthesize a negative example by randomly coupling a term that is not in $\delta$ and using all the terms in $\delta$ to form its local context. The prediction target for such a negative example is set to be the complement of a document representation of $\delta$; i.e., the complement of $\boldsymbol{BoW}(\delta)$, denoted by $\overline{\boldsymbol{BoW}(\delta)}$, which is achieved by flipping all the binary entries of $\boldsymbol{BoW}(\delta)$. To avoid confusion, hereinafter, we refer to those examples generated from training documents as *positive examples*. In our learning, we use all positive examples resulting from a training document and synthesize the same number of randomly selected negative examples for a balanced learning. Thus, for any example $k$, its input is $\boldsymbol{x}_k(\tau, \delta) = (\boldsymbol{t}(\tau), \boldsymbol{l}(\tau|\delta))$ and the learning target is the document representation of $\delta$, i.e., $\boldsymbol{y}_k(\boldsymbol{x}_k(\tau, \delta)) = \boldsymbol{BoW}(\delta)$ if $\boldsymbol{x}_k(\tau, \delta)$ is a positive example or $\boldsymbol{y}_k(\boldsymbol{x}_k(\tau, \delta)) = \overline{\boldsymbol{BoW}(\delta)}$ otherwise.

#### 4.3.2 Prediction Learning

To learn the prediction in the first stage, a subnetwork is initialized with the well-known unsupervised greedy layer-wise learning procedure with the sparse auto-encoder as its building block (Bengio et al., 2007). After a subnetwork of $H-1$ hidden layers is initialized, the subnetwork is fine-tuned by applying the document representation labels. The learning algorithm for sparse autoencoder can be found in the Appendix.

The binary nature of the output makes the cross-entropy loss suitable for this task. Given the entire training data set $(X, Y)$ of $K$ examples generated from $|\Delta|$ documents and a vocabulary of $|\Gamma|$ terms, the initial prediction loss should be defined by

$$\mathcal{L}_P(X, Y; \Theta) = -\frac{1}{2K|\Gamma|} \sum_{k=1}^{K} \sum_{j=1}^{|\Gamma|} \big((1 + y_k[j]) \log(1 + \hat{y}_k[j]) + (1 - y_k[j]) \log(1 - \hat{y}_k[j])\big),$$

where $\Theta$ is the collective notation of all the weight and bias parameters in the subnetwork. During learning, however, the sparse nature of the BoW representation often skews the target labels towards an incorrect estimation signifying that all the terms are absent in a document. To avoid a trivial solution where all terms are predicted to be absent, we re-weight the cost of a false negative error for example $k$ by $\kappa_k = |\{j: y_k[j] = 1\}_{j=1}^{|\Gamma|}|/|\Gamma|$. Thus, the actual prediction loss used in our learning is

$$\mathcal{L}_P(X, Y; \Theta) = \frac{-1}{2K|\Gamma|} \sum_{k=1}^{K} \sum_{j=1}^{|\Gamma|} \big(\kappa_k (1 + y_k[j]) \log(1 + \hat{y}_k[j]) + (1 - \kappa_k)(1 - y_k[j]) \log(1 - \hat{y}_k[j])\big). \quad (5)$$

Solving the optimization problem defined in Equation 5 based on a training data set leads to a trained deep neural network that can predict all the accompany terms in a document from their term and local context representations. Details of this learning algorithm can be found in the Appendix.

### 4.3.3 Distance Learning

After completing the prediction learning with a single subnetwork, we train a Siamese architecture by coupling two copies of the trained prediction learning subnetwork (c.f. Figure 2). When presenting a pair of input vectors $x^{(1)}, x^{(2)}$ to the two coupled subnetworks, the embedding similarity is measured by the Euclidean distance between their $CE$ representations:

$$E(x^{(1)}, x^{(2)}) = \|CE(x^{(1)}) - CE(x^{(2)})\|_2.$$

In general, the Siamese architecture needs to be trained to enhance the proper distance between term-to-term concepts conveyed in terms along with their local contexts while conserving as much of the prediction ability in each component subnetwork as possible. For example, two different terms used to describe the same concept should share the identical local context and hence their $CE$ representations should have a zero distance. However, a specific term under two different local contexts may be conveying two different concepts and hence two $CE$ representations of that term should be distant in the CE space. Measuring the local contextual similarity can be done by the LDA via the Kullback–Leibler (KL) divergence (see Section 3.2 for details):

$$KL(x^{(1)}, x^{(2)}) = \sum_{c=1}^{|\Phi|} \left( (l^{(1)}[c] - l^{(2)}[c]) \log \left( \frac{l^{(1)}[c]}{l^{(2)}[c]} \right) \right).$$

To learn the proper distance between different concepts, we take all possible situations regarding a pair of training examples into account and discover a number of objectives that contribute to the loss function used in the distance learning of our Siamese architecture. Let the binary variables $I_1, I_2$ and $I_3$ indicate three possible situations for a given input pair $(x^{(1)}, x^{(2)})$:

- $I_1 = 1$: both $x^{(1)}$ and $x^{(2)}$ are positive examples generated from a training collection or collections. In this situation, the proper distance between their representations must be learned to reflect the conceptual similarity between their local contexts. Whenever both share the same local context, in particular, their representations should be co-located or as close as possible in the CE space.

- $I_2 = 1$: both $x^{(1)}$ and $x^{(2)}$ are negative examples synthesized with a training vocabulary. In this situation, the same should be done as described in the situation corresponding to $I_1 = 1$. As the concepts conveyed in negative examples are randomly synthesized, the distance between their **CE** representations is less important than that between two positive concepts. This difference will be reflected by a treatment of weighting the similarity cost $E(x^{(1)}, x^{(2)})$ differently for $I_1 = 1$ and $I_2 = 1$ in our loss function.

- $I_3 = 1$: for $x^{(1)}$ and $x^{(2)}$ are one positive and the other is negative. In this situation, it is impossible to achieve the accurate distance between their **CE** representations given the fact that a negative example is randomly synthesized and no coherence should be possessed by such "concepts". As a consequence, their **CE** representations need to be distant as far as possible when there are similar local contexts in $x^{(1)}$ and $x^{(2)}$, which is implemented with a weighting scheme presented below.

Given two subsets $X^{(1)}$ and $X^{(2)}$ of the same cardinality $N$ of randomly selected examples from the training data set $X$ via random pairing. For a pair $n$, we denote $\mathbb{E} = E(x_n^{(1)}, x_n^{(2)})$, $\mathbb{d} = KL(x_n^{(1)}, x_n^{(2)})$ and $\mathbb{D} = e^{-\frac{\lambda}{2}\mathbb{d}}$ where $\lambda$ is a sensitivity parameter that affects the degree to which the embedding is dominated by the context similarity. We define our Siamese loss based on three situations indicated by $I_1, I_2$ and $I_3$ as follows:

$$\mathcal{L}_S(X^{(1)}, X^{(2)}; \Theta) = \sum_{n=1}^{N} \left( I_1(\mathbb{E} - \beta(1 - \mathbb{D}))^2 + I_2\rho(\mathbb{E} - \beta(1 - \mathbb{D}))^2 + I_3(\mathbb{E} - \beta)^2 \mathbb{D} \right). \tag{6}$$

Here, $\beta$ is a scaling parameter used to ensure controlled concepts' spread over the embedding space and $\rho$ is an importance parameter that weights down the importance of the situation corresponding to $I_2 = 1$. In Equation 6, three objectives work alternately with different types of training pairs specified by $I_i = 1$ ($i = 1, 2, 3$).

By combining losses defined in Equations 5 and 6, the multi-objective loss function used in the distance learning of our Siamese architecture is

$$\mathcal{L}(X^{(1)}, X^{(2)}, Y^{(1)}, Y^{(2)}; \Theta) = \sum_{i=1}^{2} \mathcal{L}_P(X^{(i)}, Y^{(i)}; \Theta_i) + \alpha \mathcal{L}_S(X^{(1)}, X^{(2)}; \Theta), \tag{7}$$

where $\alpha$ is a parameter used to balance the prediction and the distance losses applied to individual coupled component subnetworks and $\Theta_i$ is a collective notation of all the parameters in subnetwork $i$.

In our two-stage learning, all the parameters are estimated iteratively via the stochastic back-propagation (SBP) (Bottou, 2012) by optimizing the loss functions specified in Equations 5 and 7. The optimal hyper-parameters are found via a grid search with cross-validation and, in general, an early stopping criterion is applied in the SBP learning (see the next section for our specific experimental setting and the Appendix for a generic setting). In each iteration of the SBP, a small batch of training examples are randomly selected to update the parameters in training either single subnetwork (for the prediction learning) or Siamese architecture (for the distance learning). It is also worth stating that the two subnetworks in Siamese architecture are always made identical via averaging their weights and biases after each iteration of the distance learning. Details of our learning algorithms and their derivation can be found in the Appendix.

## 4.4 OOV Term Contextualized Embedding

Upon applying the contextualized semantic representation learned from co-occurring terms, the issue of *out-of-vocabulary* (OOV) term contextualized embedding has to be addressed. In general, a test document may contain more than one OOV terms. As our learning model relies on the LDA in generating the local context representation but the LDA does not address the OOV issue, we always use only those in-vocabulary terms appearing in this test document when generating the LDA-based local context representation. Without loss of generality, we need to take a document containing only one OOV term into account. Based on the learning model and the resultant contextualized semantic representation, we propose two methods to deal with the OOV situation. Let $\delta = \{\tau_{oov}, \delta_{iv}\}$ denote a test document of an OOV term $\tau_{oov}$ where $\delta_{iv} = \{\tau_i\}_{i=1}^{m}$ is a collective notation of $m$ in-vocabulary terms in $\delta$.

Our first method relies on the representation capacity of the term representation described in 3.1. We can extend the term representation to an OOV term $\tau_{oov}$. For $\tau_{oov}$, its $tfidf$ values are measured the same as done for in-vocabulary terms with all the training documents plus all test documents containing it. However, the aggregation is only done against all the in-vocabulary terms to achieve $t(\tau_{oov})$. This extension ensures that the same number of features is used to represent both in-vocabulary and OOV terms. In addition, each feature always means relatedness against the corresponding term. Similarly, we can achieve its local LDA-based context representation $l(\tau_{oov}|\delta_{iv})$ by using $\delta_{iv}$. Thus, feeding $x(\tau_{oov}, \delta_{iv}) = (t(\tau_{oov}), l(\tau_{oov}|\delta_{iv}))$ to a trained subnetwork leads to its contextualized semantic representation $CE(x(\tau_{oov}, \delta_{iv}))$. As OOV terms were not seen in training, a document of any OOV term $\delta = \{\tau_{oov}, \delta_{iv}\}$ is actually equivalent to the settings of negative examples (c.f. Section 4.3.1). In our distance learning, we have made all negative examples distant from positive examples in the CE space as far as possible. When measuring the contextualized term-to-term relatedness between an OOV term and other in-vocabulary terms, we stipulate that its most related in-vocabulary term is the one furthest in distance in the CE space. As this treatment relies on the term representation, we name it the *feature-based* OOV method.

Our second method is motivated by the coherent nature of co-occurring terms in a document and the capacity of our contextualized semantic representation in encoding term-level and document-level semantics. As a result, the contextualized semantic representation $CE(\tau_{oov}| \delta_{iv})$ should be co-located with $\{CE(x(\tau_i, \delta_{iv}))\}_{i=1}^{m}$ in the CE space and shares the group-level semantics represented in the CE space. Thus, we directly define the contextualized semantic representation of $\tau_{oov}$:

$$CE(\tau_{oov}| \delta_{iv}) = \frac{1}{m}\sum_{i=1}^{m} CE(x(\tau_i, \delta_{iv})).$$

Intuitively, we treat the OOV term as missing data and then use the centroid of $CE$ representations of $m$ co-occurring in-vocabulary terms to represent the concept defined by this OOV them along with the shared local context. As this process does not involve the OOV term features, we deliberately use the notation $CE(\tau_{oov}| \delta_{iv})$ to distinguish from the OOV term representation $CE(x(\tau_{oov}, \delta_{iv}))$ achieved by the first method. As this treatment is based on the $CE$ representations of those accompany in-vocabulary terms in the document containing OOV terms, we dub it the *concept-based* OOV method.

# 5 Experiments on Concept Embedding Learning

In this section, we describe the experimental settings and visualize results regarding the use of our Siamese architecture to learn the concept embedding, CE space, from a number of corpora in different domains. We first describe all the corpora used in our work including the semantic priming application presented in the next section. Then we give the details of our experimenting settings in the CE learning. Finally, we visualize the contextualized semantic representations of terms learnt by our Siamese architecture to demonstrate the main properties underling the *CE* representations.

## 5.1 Data sets

In our experiments, we employ six publically accessible corpora of multi-term documents from two domains: textually tagged music and multi-labeled images. Those data sets are often used as benchmarks in the two domains for different information processing tasks including MMIR.

Three music tagged corpora are CAL500, MagTag5K and Million Song Data set as follows:

- **CAL500** (Turnbull et al., 2007) is collected through public survey. Each document was annotated by multiple human participants using a controlled vocabulary. The appropriateness of each tag to each song was determined via majority vote. The corpus has a vocabulary of 158 different tags and contains 500 unique annotation documents where the cardinality or number of tags in a document is 25 on average.
- **MagTag5K** (Marques et al., 2011) is a controlled version of the MagnaTune[2] data set by removing all repeated or contradicting tags. MagnaTune is collected via an online annotation game, called TagATune, which asked two players to describe the music they were individually listening to and shared the descriptions from one to the other. Then, the players had to decide whether they were listening to the same music or not. In this scenario, the players evaluate the appropriateness of complete tag sets for pieces of music rather than just one tag at a time (Law et al., 2009). Thus, the data set conveys reliable tagging information and quality domain-specific semantics. In this data set, there are 136 unique tags in its vocabulary and a total of 5,259 different annotation documents with cardinality five on average.
- **Million Song Dataset** (Bertin-mahieux et al., 2011) contains acoustic features and meta data of a million songs. Many of the songs are annotated with tags collected from Last.fm[3], a crowd-sourced music recommendation website. As a result, the corpus has a vocabulary of 24,499 unique tags and contains 218,754 tagged documents with average document cardinality of 8.5.

As illustrated in Figure 3, the three music tagging data sets exhibits different yet typical aspects of music tagging corpora. It is observed that CAL500 and MagTag5K represent two different music tagging styles. The tag-usage distribution in MagTag5K has a long tail, i.e., most of the tags are rarely used, while CAL500 is less affected by this issue due to its use of a controlled vocabulary. A music piece was annotated by many tags in CAL500 while a much smaller number of tags were used per document in MagTag5K. In contrast, Million Song Dataset (MSD) has a huge tag vocabulary and a large number of documents. Applying our learning model to such a large data set demands additional techniques and more powerful computing resources as will be discussed later on. In our experiments, we used only CAL500 and MagTag5K, respectively, as our training corpora to investigate the influence of different tagging styles in our CE learning and the MSD as a test data set to examine the generalization of contextualized semantics learnt from a specific data set via cross-corpora setting as described in Section 5.2. As there are 75 common tags in MagTag5K and MSD, we use the Siamese architecture trained on MagTag5K in the cross-corpora test although the statistics underlying tagging is considerably different between MagTag5K and the MSD as demonstrated in Figure 3(b) and 3(c). In the MSD, we found a subset of 817 annotation documents consisting of only MagTag5K tags. Figure 3(c) shows the statistics underlying this subset. It should be noted that while tags in CAL500 and MagTag5K are ordered in descending usage order, MSD tags were reordered to match their IDs in MagTag5K for this statistic visualization only. In the MSD, there are also 39,507 documents consisting of both in-vocabulary tags and at least one OOV tag in terms of the MagTag5K vocabulary. This larger subset of the MSD is also used in the cross-corpora test to evaluate our proposed OOV methods as presented in the next section.

---

[2] https://magnatune.com/

[3] http://www.last.fm/api

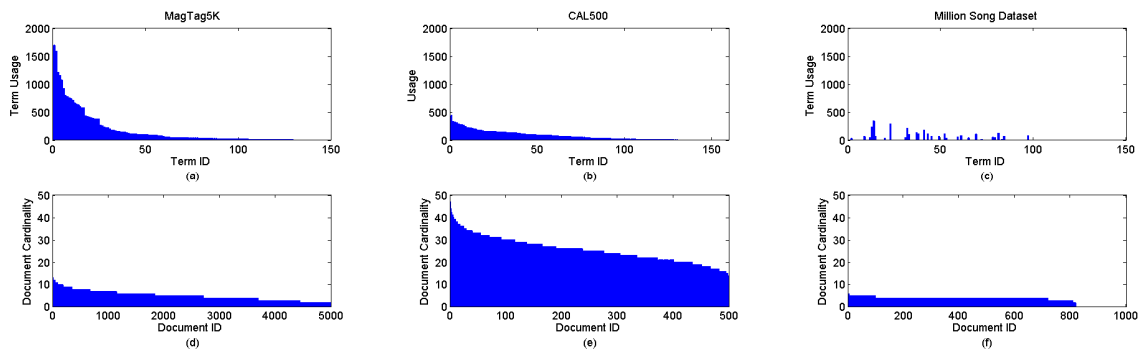

Figure 3: Statistics on three music tagging datasets. (a-c) Tag-use distribution. (d-f) Document cardinality distribution.

In addition to the music tagging data sets, three multi-labeled image data sets used in our experiments are Corel5K, LabelMe and SUNDatabase described as follows:

- **Corel5K** (Duygulu et al., 2002) was manually annotated by experts who assigned up to five labels to the most prominent objects in each image. In this corpus, images were annotated with 292 unique labels and there are 4,524 different annotation documents of 3.5 labels on average. Although all the images themselves are not accessible due to copyright, their labels are publically available, which meets our learning requirements.
- **LabelMe data set** (Russell et al., 2007) contains images labeled by crowed-sourcing via an online labeling tool. This data set contains images where all the objects were completely annotated with multiple labels while most images were only partially annotated for main or focused objects. In the corpus, images were annotated with 2,385 unique labels and there are 26,945 annotation documents containing 7.3 labels per document on average. Both images and their annotations are publically accessible.
- **SUNDatabase benchmark** (Xiao et al., 2010) was created in several stages. First, place names such as airport, bridge and city were collected from WordNet. Then, an online web search was undertaken in order to collect the categorized images in terms of different places names. Finally, the labels were refined manually. In this benchmark data set, images were annotated by 1,908 unique labels and there are 23,743 different annotation documents of 11 labels per image on average.

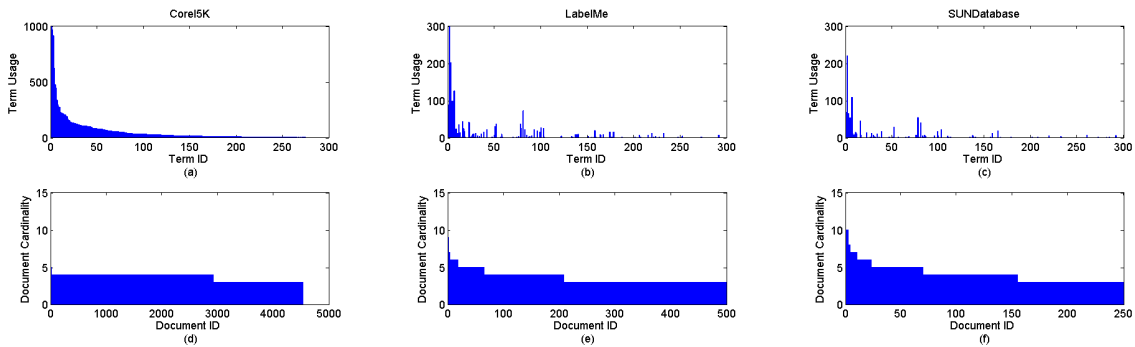

Figure 4: Statistics on three image labeling datasets. (a-c) Label-use distribution. (d-f) Document cardinality distribution.

Figure 4 illustrates the label usage and document cardinality distributions underlying those documents used in our experiments regarding the three image corpora. It is observed from Figure 4 that the statistics underlying the annotations appears to be somewhat similar for the three image corpora. In particular, there exists similar usage statistics including label-use frequency and document cardinality in both Corel5K and LabelMe. The statistical similarity underlying different corpora suggests that annotators have larger agreement on the meaning of visual concepts than that of music concepts as shown in Figure 3. For the same reason as described above, we used only Corel5K as our training corpora in our CE learning and LabelMe and SUNDatabase as test data sets in the cross-corpora setting. In the LabelMe data set, there are a subset of 520 annotation documents consisting of only Corel5K labels and a larger subset of 8,703 documents consisting of both in-vocabulary labels and at least one OOV label in terms of Corel5K. From the SUNDatabase corpus, we achieve a subset of 266 annotation documents consisting of only Corel5K labels and a larger subset of 11,935 documents consisting of both in-vocabulary labels and at least one OOV label in terms of Corel5K. Only those subsets of LabelMe and SUNDatatset are used in our experiments and their statistics are shown in Figure 4(b) and 4(c), respectively.

## 5.2 Experimental Settings

Now, we describe the experimental settings in training our Siamese architecture on three training corpora: CAL500, MagTag5K and Corel5K.

For feature extraction, we applied methods described in Section 3 to generate the term, the local context and the document representation from each document. We can achieve the feature vectors of any in-vocabulary term with Equation 2 and generate the representation of OOV terms in a similar way as described in Section 4.4. In our experiments, the de-correlation of features with PCA and linear scaling of each feature was applied to the term and the local context representations in order to ensure that each feature is in the range (-1, +1). By applying Equation 3, the local context features of a document were obtained based on a trained LDA working on all accompany terms in the document. To train an LDA model, we use all the documents in a training data set. The number of topics were empirically decided by using the hierarchical process as suggested in (Teh et al., 2006). As a result, we achieved three LDA models of 25, 19 and 20 topics trained on CAL500, MagTag5K and Corel5K, respectively. Each LDA model is applied to a relevant document to generate its local context representation. Note that a trained parametric LDA model is also used in generating the local context representations for those test documents in different settings. For training the Siamese architecture, the BoW representation of a training document is achieved with Equation 4.

For model selection and performance evaluation in different settings, cross validation (CV) was used. In CV, a training corpus is randomly split into two subsets A and B with a ratio 2:1; A for training and B for validation and test. For CAL500, 40 documents in B were randomly chosen and reserved for validation during training and the rest of documents in this subset were used for test. For MagTag5K, we adopted a default setting suggested by Marques et al. (2011) instead of the random split and 300 documents were randomly selected from subset B for validation while the rest of documents were reserved for test. The same setting as done for MagTag5K was applied to Corel5K. Furthermore, it should be clarified that we have exploited MagTag5K in simulating OOV situations. To do so, we randomly reserved 22 tags from the MagTag5K vocabulary. Thus, the number of in-vocabulary tags is down to 114. Accordingly, all the documents containing any of those 22 tags are removed before the CV split. Hence, the number of documents used in the aforementioned CV setting is 3,826.

To train the Siamese architecture, we randomly generate the same number of negative examples as that of positive examples in subset A by using the procedure described in Section 4.2 and append them to subset A in each CV trial. It is worth mentioning that the use of more negative than positive examples often leads to a degenerate solution that the uniform negative output in prediction is always reached regardless of any actual input. For the distance learning, as described in Section 4.3, training documents in subset A were randomly paired so that roughly equal number of paired examples was generated for two situations corresponding to $I_i = 1$ ($i = 1, 2$) as described in Section 4.3. Consequently, the number of examples for $I_3 = 1$ doubles that number.

In the SBP learning, the "optimal" hyper-parameter values were found with a grid search during multiple CV trials and summarized as follows: a) for the sparse autoencoder learning, the sparsity factor is 2, weight decay is 0.02 and a quasi-Newton algorithm was employed for training (see Appendix for details); b) for the prediction learning, the learning rates were initially set to $10^{-4}$, $10^{-5}$ and $10^{-5}$ for MagTag5K, CAL500 and Corel5K, respectively, and then decayed with a factor of 0.95 every 10 epochs. The re-weighting parameter $\kappa$ in Equation 5 is automatically obtained for each example as described in Section 4.3.2; and c) for the distance learning, the importance and the scaling factors in Equation 6 were set to $\rho = 0.5$ and $\beta = \sqrt{d}$, respectively, and the trade-off factors in Equation 7 were set to $\alpha = 2000$ for MagTag5K and $\alpha = 1000$ for both CAL500 and Corel5K. The same learning rates used in the prediction learning were used for distance learning but the decay rule was applied every 10 mini-batches in SBP.

Early stopping principle was applied in both the prediction and the distance learning stages for generalization. Instead of monitoring only the cost defined in the loss functions on a validation set, however, our stopping criterion makes use of a surrogate loss on the validation set; i.e., the performance of a semantic priming task, $P@2$, to be described in the next Section. The motivation behind this stopping criterion comes from the unsupervised learning nature of our CE learning; the loss functions were formulated for generic semantics without ground-truth. As a generic information retrieval task, semantic priming allows us to see "ground-truth" to some extent. Hence, the actual generalization performance can be guaranteed at least on the generic semantic priming task. As a result, our stopping criterion is as follows: we evaluated the priming performance based on the representations obtained after every 200 epochs and examined the performance improvement on both training and validation data sets between two adjacent tests. The learning was stopped at the point of the smallest improvement between two test points by human inspection. We believe that this is a generic stopping criterion applicable to any applications of our contextualized semantic representation. In Section 5.3, we demonstrate that this early stopping criterion working on the semantic priming task actually leads to satisfactory concept embedding.

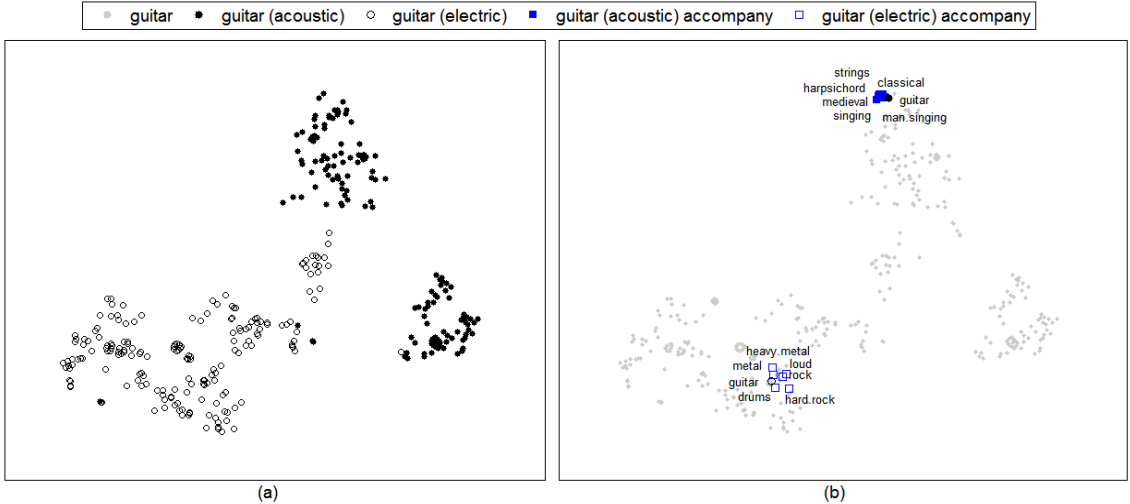

Figure 5: Visualization of the ***CE*** representations of "guitar" in MagTag5K. (a) 2-D projections corresponding to all the 388 "guitar" instances. (b) 2-D projections of the instances corresponding to the accompany tags co-occurring with "guitar" in two randomly selected documents containing "guitar".

For model selection, we examined a number of feed-forward neural networks that have hidden layers ranging from one to four layers and different numbers of hidden units in a hidden layer ranging from 10 to 200. For reliability, we always repeated the aforementioned CV experiments for three trials. As a result, the "optimal" subnetwork has a structure: $input \to 100 \to 100 \to \mathbf{\underline{10}} \to output$; i.e., a multi-layered perceptron has three hidden layers of 100, 100 and 10 hidden units where the dimension of the ***CE*** representation is 10. It is worth mentioning that our model selection described above was mainly done based on the MagTag5K training set. For training on CAL500 and Corel5K, the grid search score is much smaller thanks to the information acquired from the MagTag5K training. Actually, the optimal structure achieved based on the MagTag5K training turns out to be the best for both CAL500 and Corel5K as well. Hereinafter, we report experimental results based on this optimal structure.

**5.3 Visualization of Concept Embedding**

After the completion of the two-stage learning, a trained subnetwork provides a 10-dimensional ***CE*** representation for any given term along with its local context. By employing the unsupervised t-SNE (Maaten & Hinton, 2008), we can visualize the ***CE*** representations learnt from the training corpora by projecting the 10-dimensional representation to a 2-dimensional space. Thanks to the powerful non-linear dimensionality reduction capacity of the t-SNE, we anticipate that the visualization would demonstrate the main properties of contextual semantics and relatedness learnt from training corpora in different domains vividly.

First of all, we choose "guitar" to be the focused tag as it is a typical example of a tag that can convey different concepts in the presence of different local contexts (c.f. Section 1). We collect all the 388 documents containing "guitar" from the MagTag5K data set and apply the subnetwork trained on MagTag5K to produce their ***CE*** representations for all 388 "guitar" with different local contexts. To facilitate our presentation, hereinafter, *instance* is used to describe an embedding vector of a focused term along with its local context. Figure 5 shows the projection of all the ***CE*** representations of 388 "guitar" instances onto a 2-D space as well as the projection of ***CE*** representations of some relevant tags that share the same local context with the focused tag. It is observed from Figure 5(a) that the concepts defined by the tag "guitar" instances are grouped into three clusters, which demonstrates that our *CE* representation captures multiple meanings of "guitar" in different contexts. By a closer look, we find that three clusters actually correspond to two different meanings or concepts: "acoustic guitar" indicated by the solid circle (●) and "electric guitar" indicated by the hollow circle (○). As two different instruments are often used in different music genres, our *CE* representation has successfully distinguished between them by embedding them in different regions in the CE space. Figure 5 (b) further shows the projection of *CE* representations corresponding to the accompany tags from two randomly chosen documents containing "guitar", one indicated by solid square (■) from an "acoustic" cluster and the other indicated by hollow square (□) from the "electric" cluster, by superimposing them on the projection of "guitar" as shown in Figure 5(a). Note that we deliberately shade all 388 "guitar" instance projections in Figure 5(b) for clearer visualization. It is clearly seen from Figure 5(b) that different concepts in the same context have been properly co-located with each other in the CE space thanks to the distance learning used in training the Siamese architecture. Furthermore, the *CE* representations of all the co-occurring tags in a document collectively provide an alternative document-level representation reflecting a set of similar concepts and their subtle differences. While such concepts and their subtle differences in context seem to be easily grasped by people, we emphasize that the embedding shown in Figure 5 was acquired via unsupervised learning.

Next, we take the label "house" in the image domain as an example to examine whether concepts reflecting the ambient environment specified in its local contexts can be captured by our *CE* representation. Moreover, we would demonstrate the transferability of learnt contextualized semantics via visualization. As a result, we collect all the documents containing the label "house" in Corel5K, LabelMe and SUNDatabase data sets and use the Siamese architecture trained on Corel5K to generate the *CE* representations for 152 "house" instances. Figure 6 illustrates the projections of all 152 "house" instances in a 2-D space where 96 instances indicated by solid circle (●) from Corel5K, 14 instances indicated by gray circle (●) from SUNDatabase, and 42 instances indicated by hollow circle (○) from LabelMe. Due to the unavailability of images in Corel5K, we have to examine the embedding by inspecting all the relevant documents manually. In general, our inspection shows that the contextualized semantics learnt from Corel5K properly reflects concepts corresponding to different ambient environments for the "house" instances in Corel5K, LabelMe and SUNDatabase and the 2-D projections of their *CE* representations are illustrated in Figure 6. Fortunately, we can use images available from LabelMe to confirm our inspection. As a result, we present 14 "house" images in Figure 6(a) and the corresponding annotation documents in Figure 6(b). It is evident from Figure 6 that houses with similar ambient environments are close to each other in the CE space. In particular, it is observed that a manifold appears in the 2-D space and shows the transition of ambient environments from castles, seaside and rural houses to urban houses. As illustrated in Figure 6(a), we highlight the manifold by connecting those projection points on the "house" manifold in response to the ambient environmental changes. Once again, we highlight that the concepts are captured solely from the co-occurring labels in the textual form via unsupervised learning without using any image features or visual similarity.

Finally, we demonstrate OOV term embedding via visualization. As described in Section 4.4, the concept-based OOV embedding method entirely relies on the representations of in-vocabulary terms appearing in the local context and the *CE* representation of an OOV term is actually the centroid of its co-occurring in-vocabulary term representations in the CE space. One easily imagines such an embedding. As a result, we simply visualize the *CE* representation of an OOV term achieved by the feature-based embedding method (c.f. Section 4.4). In our experimental

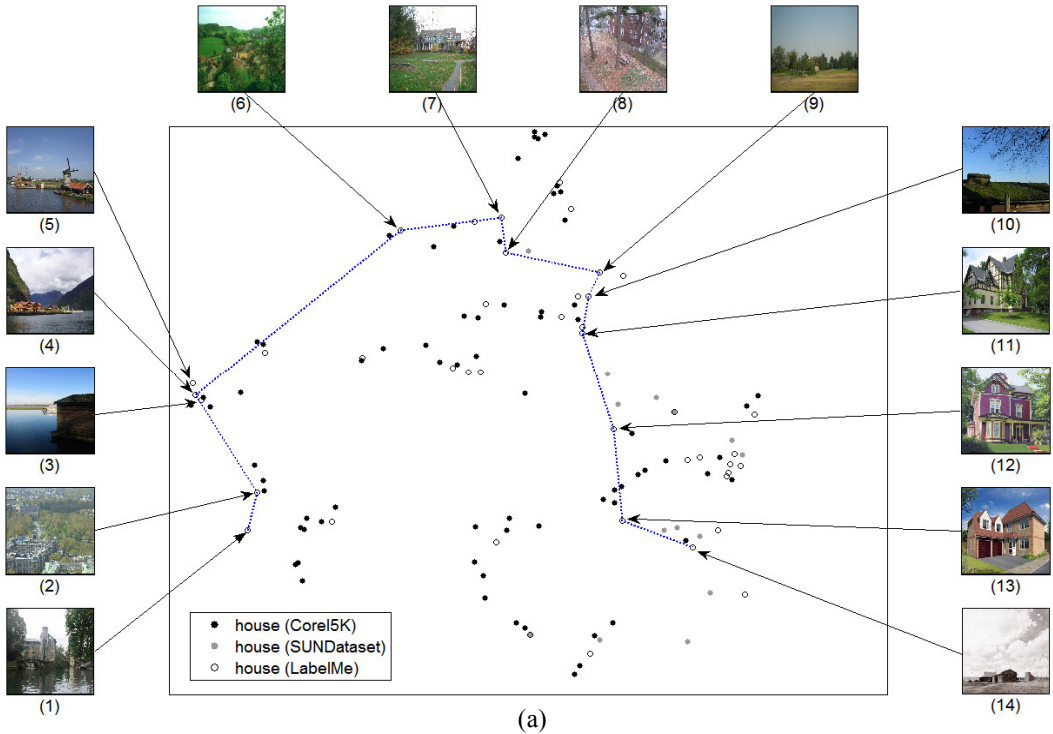

| (1) | bridge, **house**, tower | (8) | tree, grass, **house**, window, bench |
|---|---|---|---|
| (2) | city, bridge, river, **house**, castle, park, tower | (9) | sky, tree, grass, fence, **house** |
| (3) | sky, water, **house** | (10) | sky, tree, **house** |
| (4) | mountain, sky, water, **house** | (11) | sky, tree, grass, **house**, road |
| (5) | sky, water, **house** | (12) | city, sky, tree, **house**, path, window, door, garden, lawn |
| (6) | tree, forest, field, **house** | (13) | sky, grass, **house**, window, street, door |
| (7) | tree, grass, **house** | (14) | sky, land, ground, **house** |

(b)

Figure 6: Visualization of the **CE** representations of the label "house" in image datasets. (a) 2-D projections of all 152 "house" instances along with the associated images of "house" instances in LabelMe whose projections are roughly located on a manifold as indicated by connected points. (b) Annotation documents used as local contexts for those instances on the manifold.

settings described in Section 5.2, 22 tags in MagTag5K were reserved to simulate OOV terms. For visualization, we choose a typical document of four tags {"classical", "violins", "strings", "cello"} that annotates the song "La Reveuse" composed by Martin Marais. In this document, "cello" is one of 22 OOV terms. To facilitate our presentation of the main properties of an OOV term in the CE space, we also visualize all the instances derived from the incomplete document of {"classical", "violins", "strings"} including all the positive/negative examples. Consequently, the incomplete document leads to three positive and 111 negative instances by coupling all the remaining 111 in-vocabulary tags with this incomplete document (c.f. Section 4.3.1). Hence, tags "classical", "violins" and "strings" are in turn to be the focused tags in three positive instances and the document containing this tags and {"classical", "violins" and "strings"} together form its local context. To generate a negative instance, we substitute the focused tag in the positive instance with an in-vocabulary tag other than "classical", "violins" and "strings". Figure 7 illustrates 2-D projections of the **CE** representations of "cello" and all relevant instances specified above. It shows the projections of all the instances concerning the exemplar document as described above. It is observed from Figure 7 that all three positive instances indicated by blue square (■) are co-located and projected onto a tiny region at the upper right corner of the 2-D space. With the music knowledge, we see that all three instances correspond to concepts that classical music is played by string instruments. In contrast, 111 negative instances indicated by cross in red (×) and are projected to two regions in the 2-D space; i.e., the small region consisting of seven instances is close to three positive instances and the large one composed of the remains negative instances is far from the small region as well as those projections of three positive instances as shown in Figure 7. A closer look at those near three positive instances reveals that the tags used to form those instances, as depicted in Figure 7, are actually semantically associated with the positive instances (even though they are treated as negative). Any of those tags might have been used to annotate this music piece without altering the concept; i.e., most classical string-based music is orchestral in an old style, probably from the Baroque era and rarely involving piano in it. Moreover, most such music includes the fiddle as instrument. This result demonstrates the capability of our approach in capturing the accurate concepts underlying training documents even for those treated as "negative" in training. From Figure 7, it is seen that the OOV tag instance ($\tau_{oov}$ ="cello" and $\delta_{iv}$ ={"classical", "violins", "strings"}) indicated by green star (★) is projected into the large region of negative example due to the fact that an OOV term was not seen in training and hence the resultant OOV instance has to be treated as negative (c.f. Section 4.4). As the OOV instance is further from the three points corresponding to three positive instances than any negative instances in the 2-D space, the visualization in Figure 7 also intuitively provides the evidence to support our measure of the contextualized term-to-term relatedness between in-vocabulary and OOV terms in the feature-based OOV treatment.

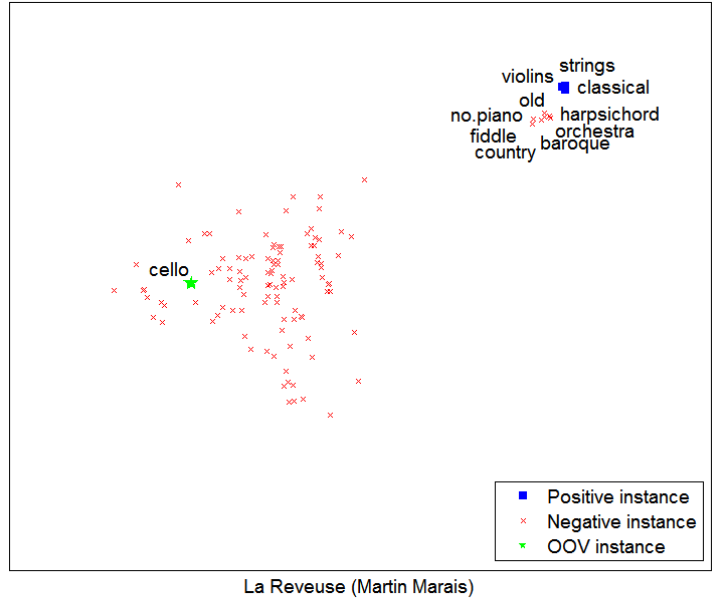

Figure 7: 2-D projections of the **CE** representation of the OOV tag "cello" along with those sharing the same local context.

In summary, visualization shown above suggests that our leaning model successfully captures contextualized semantics from co-occurring terms in different domains and also demonstrates its capability in dealing with domain-specific semantics, transferability of learnt semantics across different corpora and the OOV terms. In addition, visualization also suggests that the use of a surrogate loss, i.e., semantic priming performance, in our stopping criterion during learning leads to generic **CE** representations applicable to various tasks described in Section 1.

# 6  Application to Semantic Priming

As demonstrated in Section 5, our trained model captures high quality terms semantics that tends to be generic and hence can support a variety of applications. Semantic priming is an application that depends directly on those semantics without the need of accessing content or other information regarding media (Lund & Burgess, 1996; Osgood, 1952). As priming highlights the versatility of the semantics from an abstract point of view, it provides invaluable insight into the performance of a semantic model regardless of different applications. Hence, we employ this generic task to evaluate the performance of our proposed approach based on those data sets described in Section 5.1 and further compare ours to a number of state-of-the-art methods on learning semantics from co-occurring terms for thorough evaluation.

## 6.1  Semantic Priming and Evaluation

In general, semantic priming is a process involving associating concepts based on their semantic relatedness. This abstract process is often used to evaluate the learnt semantics and demonstrate the performance of a semantic learning model (Lund & Burgess, 1996). Ideally, coherent terms should be associated properly with one another based on the intrinsic contextualized semantics conveyed by them. To do so, a semantic learning model has to resolve the highly nonlinear relationship between terms and contexts by capturing intentions behind those observed terms as accurately as possible. Thus, the semantic priming task becomes a proper test bed to evaluate the capabilities of a semantic learning model by measuring the relatedness of terms in different scenarios such as applicability to new documents, incomplete context and the presence of OOV terms.

Below, we first present the priming protocol used in evaluating a term-based contextualized semantic representation. Then, we extend this protocol to the document-level so that all the existing semantic learning models can be compared fairly on the exact same condition. Finally, we describe the evaluation criteria used in semantic priming.

### 6.1.1  Priming Protocol

Semantic priming was first introduced by Meyer and Schvaneveldt (1971) to associate semantically related concepts to each other, e.g., doctor-nurse. Through semantic priming, it has been shown that human subjects read consecutive words quicker when the words are semantically or syntactically associated. Given a priming concept or a query concept, all other concepts can be generally split into two groups, related and unrelated concepts, depending on the context. Thus, semantic priming functions as a highly generic evaluation method for learnt semantics without access to information other than the learnt semantics themselves. Semantic priming was first used in (Lund & Burgess, 1996) in evaluating the appropriateness of learnt *similarity* where a word embedding space was built using aggregation of a textual corpus. Using such embedding, word similarity was estimated using priming such that the closer two words' representations were, the more similar they were considered to be. However, there exists a subtle difference between syntactic and semantic relatedness (c.f. Section 2.2) and we focus on the semantic relatedness in our experiments reported in this section. As a result, we define the priming as the capability of a semantic learning model in identifying related concepts given a single query concept represented as a term and its local context as defined in Section 1.

Priming is reflected by a learnt semantic representation where similarity between concepts is encoded in some semantic distance. In contextualized semantics, such distance is significantly affected by the context. A contextualized semantic model uses the context to express concepts meaningfully via their contextualized semantic distance denoted by $\mathbb{e}(\tau_1, \tau_2 | \delta)$ where $\tau_1$ and $\tau_2$ are two different terms in the document $\delta$ that forms their shared local context. This distance measure can straightforwardly be applied to priming over a set of concepts as follows: given a query concept $\tau$, all available terms $\tau_i$ ($i = 1, ..., |\Gamma|$) in a vocabulary $\Gamma$ are ranked based on their corresponding contextualized semantic distances to the query concept:

$$Prime(\tau, \delta) = \left(\tau_i \mid \forall \tau_i, \tau_j \in \Gamma \colon \mathbb{e}(\tau, \tau_i | \delta) \leq \mathbb{e}(\tau, \tau_j | \delta) \; if \; i \leq j\right)_{i=1}^{|\Gamma|}. \tag{8}$$

Intuitively, Equation 8 results in an ordered list of all $|\Gamma|$ terms whose corresponding representations have increasing distances away from the query concept. Ideally, the terms of contextualized semantic similarity to the query concept should precede those sharing no such semantic similarity, and the top term of the list may correspond to the query concept itself.

For a semantic learning model, acquiring such ranked list for a specific query concept depends only on the definition of a distance metric used in its semantic representation space. In literature, the priming performance evaluation requires ground-truth or gold standard of all the different concept similarities in terms of all possible contexts. Unfortunately, such information is not only missing for descriptive terms so far but also does not seem attainable in general since it demands the human judgment on terms' relatedness in an unlimited number of contexts.

To alleviate the problem of the ground-truth unavailability, we assume that all co-occurring terms in a single document are coherent and hence, semantically similar in terms of their shared context. Thus, one document is used as ground-truth; each time one of its constitutional terms is used as a query term to prime other terms in that document. As a result, the priming protocol used in our experiment is as follows. Given a document, each term in this document would be used in turn as a query term that couples with the shared local context derived from the document to form a query concept for priming. The list of primed terms resulting from each query term is then compared against this document (treated as ground-truth) to measure the priming accuracy as described in Section 6.1.3. Thus, the performance of a contextualized semantic learning model is evaluated by taking the priming accuracy on all the evaluation documents into account.

### 6.1.2 Extended Priming Protocol

The priming protocol specified in Section 6.1.1 is used in evaluating a contextualized semantic learning model where the information in an entire document is required. However, there are many different semantic learning models that do not consider the local context, e.g., all the models in learning global relatedness such as PCA and LSA as reviewed in Section 2.1. It seems unfair if we compare a contextualized semantic learning model to those that work only on a single term without access to the document-level information. To allow us to compare ours to more state-of-the-art semantic learning models, we extend the priming protocol defined in Section 6.1.1 by allowing all semantic learning model to use exactly the same information conveyed in an entire document in semantic priming. Hence, any model is provided with an entire document and the collective priming results of all the terms in this query document will be used for performance evaluation. In other words, the extended priming amounts to merging all the ranked lists achieved by different terms in the query document into a single document-level global ranked list with the same distance metric in the semantic representation space. For a term in the query document, however, priming itself actually results in a zero distance situation. If this result is allowed, almost all models can yield an error-free priming result. Therefore, we have to exclude such priming results of zero distance in our extended priming protocol. Given a query document $\delta$ and a vocabulary $\Gamma$, the extended priming is defined by

$$E\_Prime(\delta) = \left(\tau_i \mid \forall \tau_i, \tau_j \in \Gamma; \; \forall \hat{t} \in \delta \wedge \hat{t} \neq \tau_i \colon \min\{\mathbb{e}(\hat{t}, \tau_i | c)\} \leq \min\{\mathbb{e}(\hat{t}, \tau_j | c)\} \; if \; i \leq j\right)_{i=1}^{|\Gamma|}. \tag{9}$$

In other words, Equation 9 results in a ranking list of $|\Gamma|$ terms by using the minimum distance between any term $\tau_i$ in $\Gamma$ and all the $|\delta|$ terms in a query document $\delta$. As this protocol is designed for any semantic learning models no matter whether it uses the context, this distance measure $\mathbb{e}(\hat{\tau}, \tau_i|c)$ is decided by the nature of a semantic learning model; i.e., $c = \delta$ for a contextualized model and $c = null$ otherwise. In Equation 9, the condition "$\forall \hat{\tau} \in \delta \wedge \hat{\tau} \neq \tau_i$" ensures that the zero distance information is never counted in finding out the minimum distance for ranking. Thus, this protocol guarantees that all semantic learning models are fairly compared by performing this document-level semantic priming with exactly the same input and exactly the same form in expressing their priming results.

### 6.1.3 Priming Accuracy

In general, the priming performance is measured by the precision at $K$ denoted by $P@K$; i.e., the precision when only the top $K$ entries in a ranked list are considered on a reasonable condition that $K$ is less than the number of in-vocabulary terms $|\Gamma|$. Here, we denote the top $K$ ($K \leq |\Gamma|$) entries in a primed list by $Prime_K(\tau, \delta) = (\tau_i)_{i=1}^{K}$ in the priming protocol (c.f. Equation 8) or $E\_Prime_K(\delta) = (\tau_i)_{i=1}^{K}$ in the extended priming protocol (c.f. Equation 9). For a document $\delta$, the priming list achieved based on a priming protocol is thus defined by

$$Prime_K(x) = \begin{cases} Prime_K(\tau, \delta) & x = (\tau, \delta) \text{ for the priming protocol} \\ E\_Prime_K(\delta) & x = \delta \text{ for the extended priming protocol} \end{cases}$$

Then $P@K$ precision is defined as the ratio of primed terms in the ground-truth (i.e., all the terms in the query document $\delta$) out of all $K$ primed terms:

$$P@K(x) = \frac{|Prime_K(x) \cap \delta|}{K}.$$

Note that this measure is applicable to a query term in the priming or a query document in the extended priming protocols. For an evaluation data set of multiple examples, $X = \{x_i\}_{i=1}^{|X|}$, the overall $P@K$ precision is defined by

$$P@K(X) = \frac{\sum_{i=1}^{|X|} P@K(x_i)}{|X|}. \tag{10}$$

Intuitively, up to a prime level $K$, $P@K$ measures the precision of primed terms against the ground-truth to find out how many related terms appear in the top $K$ primed terms. Due to the limitation of the ground-truth, the $P@K$ measure may be affected by the cardinality of a query document, i.e., $|\delta|$. In other words, only up to $|\delta|$ primed terms can be confirmed definitely with the ground-truth. As $K$ exceeds $|\delta|$, $P@K$ values might decrease rapidly for those documents of few terms. Although $P@K(X)$ may be a reasonable measure of comparison when $K \leq |\delta|$, it does not faithfully reflect the performance of any models when $K > |\delta|$. In contrast, the averaging precision on all the $P@K$ ($K = 1, \dots, |\delta|$), i.e., $AP(x) = \frac{\sum_{K=1}^{|\delta|} P@K(x)}{|\delta|}$, automatically adapts for the various lengths of different documents used as ground-truth by only concerning the top $|\delta|$ entries of the primed list resulting from a query instance, which provides a reliable performance measure. Similarly, the overall averaging precision on a test data set of $|X|$ examples is

$$MAP(X) = \frac{\sum_{i=1}^{|X|} AP(x_i)}{|X|}. \tag{11}$$

In essence, semantic priming in response to a query concept is an information retrieval task and hence the evaluation measures commonly used in information retrieval are applicable. The Area Under Curve (AUC) is a commonly used measure by calculating the area formed under the curve of precision as a function of recall at the standard 11 recall levels: $\mathbb{l} = \{0.0, \ 0.1, \dots, 1.0\}$ (Manning et al., 2008, pp. 158–163). Precision and recall at a specific recall level $\ell \in \mathbb{l}$ are defined by:

$$Precision(\ell|x) = \frac{|Prime_K(x) \ \cap \ \delta|}{K},$$

where

$$\ell = Recall(K|x) = \frac{|Prime_K(x) \ \cap \ \delta|}{|\delta|}.$$

$Recall(K|x)$ specifies a certain recall level $\ell = \frac{k}{|\delta|}$ implying that at least of $k$ out of all the $|\delta|$ related terms in the ranked list have been retrieved and is used to form the measure $Precision(\ell|x)$, i.e., precision $P@K$ at level $\ell$. Accumulating the precision values at all the 11 recall levels across an evaluation data set leads to an overall AUC measure:

$$Precision(\ell|X) = \frac{\sum_{i=1}^{|X|} Precision(\ell|x_i)}{|X|}, \ \ell = 0.0, 0.1, \dots, 1.0. \tag{12}$$

Intuitively, a larger AUC region formed by $Precision(\ell|X)$ suggests that more of the related terms have been retrieved at the standard recall levels and the precision-recall curve clearly shows the performance of a tested model at different recall levels.

In summary, four criteria, $P@K$, $MAP$, Precision vs. Recall and $AUC$, defined above are used in our experiments to evaluate the priming/extended priming performance of a semantic learning model.

### 6.2 Experimental Protocols

For a thorough performance evaluation in semantic priming, we have designed a number of experiments in different settings corresponding to several real scenarios to test the learnt semantic representations, including: **a) domain-specific semantics**: test on all the documents used in training a model and those unseen documents in the same corpus; i.e., a subset of documents were not used in training; **b) transferability**: test on the different corpora where none of documents in those corpora were used in training; **c) noisy data**: test on incomplete local context; **d) OOV data**: testing on synthesized and real documents of OOV terms; and **e) Comparison:** comparing ours to those semantic learning models reviewed in Section 2 with exactly the same settings. As described in Section 5.2, we conducted the cross-validation in training a semantic model for three trials. Hence, the averaging accuracy along with standard error arising from three trials are reported in terms of two priming protocols described above.

### 6.2.1 Within-Corpus Setting

The within-corpus setting refers to the evaluation that uses the training or the reserved test documents subsets from the training corpora (c.f. Section 5.1); i.e., CAL500, MagTag5K and Corel5K. The use of training documents in this setting is expected to test the quality of semantics learnt by a model in terms of this application. Moreover, measuring the priming accuracy on the train documents mimics a real scenario where all available information is used in building up a semantic space to be used in a variety of applications later on. On the other hand, by using the test document subset in this setting, we would evaluate the generalization of the learnt semantics into unseen documents that were probably annotated by the same cohort of users. We refer to such evaluation as *within-corpus test* (WCT) and expect that this setting would examine the quality and the generalization of learnt semantics in a domain-specific sense.

### 6.2.2 Cross-Corpora Setting

Unlike the WCT, we design experiments to test unseen documents in corpora that were never used in semantic learning. We refer to this type of evaluation as *cross-corpora test* (CCT). As a result, the CCT would investigate the transferability of learnt semantics in terms of this application. In our CCT experiments, semantic representations acquired on MagTag5K were applied to the documents in Million Song Dataset (MSD) and those acquired from Corel5K were applied to the documents in LabelMe and SUNDatabase. As described in Section 5.1, we have to limit the test documents to those that contain only the in-vocabulary terms appearing in a training corpus. As a result, there are 817, 520 and 266 eligible test documents in MSD, LabelMe and SUNDatabase, respectively.

### 6.2.3 Incomplete Local Context Setting

To achieve a contextualized semantic representation of a term, both the term and its local context, i.e., all accompany terms in the document containing it, must be required as described in our problem formulation in Section 1. In real applications, a test document could mismatch training data. For instance, it could be a subset of a training document by using fewer yet more informative terms or an enhanced version of a training document by adding more coherent terms. In this setting, we would design experiments to test mismatched documents. As argued in Section 6.1.1, it does not seem possible to attain all the different concepts and their similarities in terms of all possible contexts. Thus, it is impossible for us to simulate on one mismatch situation that more coherent terms are added to existing documents. Fortunately, we can simulate the other mismatch situation, incomplete local context, by removing a few terms from existing documents.

To simulate the incomplete local context situation, a training example of complete local context, $x = (t(\tau), l(\tau|\delta))$, is altered into a corrupted version, $\tilde{x} = (t(\tau), l(\tau|\tilde{\delta}))$ where $\tilde{\delta}$ is a subset of the original document $\delta$ achieved by removing a number of terms randomly from $\delta$. The incomplete context, $l(\tau|\tilde{\delta})$, corresponds to the topics distribution obtained from the incomplete document $\tilde{\delta}$. As a result, the use of fewer accompany terms in $\tilde{\delta}$ results in larger uncertainty in semantic priming and hence causes a bigger difficulty in priming all the accompany terms in the original document $\delta$. Here, we emphasize that the ground-truth is the original document but the local context is derived from a subset of this document in semantic priming under this setting. In our incomplete local context experiments, we used the missing rate defined by $\left(1 - \frac{|\tilde{\delta}|}{|\delta|}\right) * 100\%$ to control the number of terms removed randomly from a complete document. In this paper, we report results based on the missing rate in different ranges: up to 10%, between 10% and 30% and between 30% and 50% due to the variable length of different documents.

### 6.2.4 Out of Vocabulary (OOV) Setting

The OOV problem appears challenging in semantic learning from descriptive terms. Based on our proposed approach, we have proposed two methods to deal with OOV terms as described in Section 4.4. Here, we would use semantic priming to evaluate our proposed methods.

In our OOV experiments, we used the reserved subset of MagTag5K as described in Section 5.1. In this reserved subset, there are 1,160 documents where each of them contains at least one out of the 22 reserved terms used as simulated OOV terms. Their concept **CE** representations achieved from the semantic model trained on MagTag5K were used in semantic priming. Moreover, we also used the real documents containing OOV terms in the test corpora (c.f. our CCT setting in Section 6.2.2). As a result, there are 39,507 documents involving 23,619 OOV terms in the MSD, 8,703 documents of OOV 2,110 terms in the LabelMe and 11,935 documents containing 2,068 OOV terms in the SUNDatabase used in our OOV experiments. For those OOV documents in the MSD, the semantic model trained on MagTag5K was used to generate their **CE** representations. For those OOV documents in the SUNDatabase and the LabelMe, the semantic model trained on Corel5K was employed to yield their **CE** representations.

To the best of our knowledge, those approaches used in our comparative studies (c.f. Section 6.2.5) do not address the OOV issue. Hence, the OOV experiments only involve our proposed approach described in Section 4.4 and the priming protocol is only employed for performance evaluation.

### 6.2.5 Comparison Settings

We use the learning models reviewed in Section 2 as baselines for comparative studies. For training and test, we apply the exact same cross-validation protocol described in Section 5.2 to each semantic learning model. As a result, the information on training those models is summarized as follows:

**Latent Semantic Analysis (LSA)**: As described in Section 2, the unsupervised dimensionality reduction technique is performed using the training documents and model selection was done by using the percentage of variance (POV) measure by monitoring eigenvalues $\lambda_i$ resulting from the matrix decomposition. As a result, $n$ features are employed when the top $n$ eigenvalues cover at least 90% of the variance of training data; i.e., $POV = (\sum_{i=1}^{n} \lambda_i)/(\sum_{i=1}^{|\Gamma|} \lambda_i) \geq 90\%$. As a result, we retained 35, 25 and 80 features for MagTag5K, CAL500 and Corel5K, respectively. The same numbers of features were extracted for test data.

**Principle Component Analysis (PCA):** PCA relies on preprocessing and aggregation of the document-term binary matrix followed by dimensionality reduction of the aggregated matrix in order to obtain per term feature vectors. In our experiments, we applied preprocessing techniques including: using the binary term frequency, the $tfidf$ re-weighting and Positive Point-wise Mutual Information (PPMI) re-weighting as preprocessing. Also we considered different distance metrics in measuring the term-to-term relatedness such as the cosine, the co-occurrence (non-normalized cosine), Kullback–Leibler divergence and Hellinger divergence as aggregation measures. For each of those combinations of preprocessing and aggregation, we performed unsupervised dimensionality reduction and evaluated the resultant semantic space by using documents from MagTag5K and CAL500 based on their $P@2$ priming performance. The combination that produced the best results is the $tfidf$ reweighed matrix followed by the co-occurrence aggregation measure. As a result, we shall report results based on this combination.

**Information Theoretic Smoothing (InfoTheo):** This model started with smoothing the binary BoW representation of each document by using information regarding the pairwise use of terms over an entire training data set. Following the suggestions in (Mandel et al., 2011), we obtained this pairwise information by using all the 12 aggregation methods listed in the PCA setting and applied such information to the smooth document-term matrix generation. This matrix requires a further tuning of the two parameters, the number of associated terms $k$ and the reweight factor $\alpha$. With the suggestions in (Mandel et al., 2011), we tuned those parameters by a grid search on a reasonable range for each of training data sets with different aggregation methods, respectively. We looked into the situations as $k = 1, 3$ and $5$ terms while reweighting the matrix using different factors for $\alpha = 0.1, \ldots, 0.5$. A total of 180 experiments were carried out in each of MagTag5K and CAL500. We observed that the $tfidf$ reweighed matrix followed by the co-occurrence aggregation measure performed significantly better than other 11 aggregation methods. As a result, we applied the best aggregation method to MagTag5K, CAL500 and Corel5K. We report results based on the setting corresponding to the best $P@2$ training performance in this paper. In detail, the optimal parameters are $k = 1$, $\alpha = 0.2$ for MagTag5K, $k = 1, \alpha = 0.1$ for CAL500 and $k = 3, \alpha = 0.3$ for Corel5K. Furthermore, we conducted experiments by using a full term-to-term matrix and PCA dimensionality reduced version of this matrix. We observed that the dimensionality reduced version generally outperform the full matrix on different data sets. Also this processing allows both InfoTheo and other global relatedness learning model to have the same dimension in their representation spaces. In this paper, we report only the results generated by the dimensionality reduced version.

**Skip Gram:** In order to avoid capturing any unreal syntactic structure, we randomize the order of terms in each document before the Skip Gram learning. Training a Skip Gram model requires tuning two hyper-parameters: dimension of the embedding space and size of the neighborhood window used to specify the context. Using a grid search, we trained a number of Skip Gram models for a training data set and selected the one with the best $P@2$ training performance to report their results in this paper. We observed that the performance on different data sets was not sensitive to the dimensionality of the embedding space but affected by the window size. In general, the smaller the window size, the better the model performed. As a result, we selected the models that had the window sizes of one, three and one for MagTag5K, CAL500 and Corel5K, respectively, and the dimension of the embedding space on the three data sets is the same used for PCA, i.e., 35, 25 and 80 for MagTag5K, CAL500 and Corel5K, respectively. In our experiments, we use the word2vec source code (Mikolov et al., 2013) to train the Skip Gram models.

**Latent Direchlet Allocation (LDA) and Probabilistic LSA (PLSA):** the same number of features used in representing our local context was employed for the LDA evaluation; i.e., 19, 25 and 20 features were used for MagTag5K, CAL500 and Corel5K, respectively. The hyper-parameter tuning was described in Section 5.2. For LDA, we used the standard C implementation of LDA (Blei, Ng & Jordan, 2003) in our experiments. The same number of topics used in LDA was adopted for the PLSA as the unique difference in the two methods is the use of different distributions in capturing document-level semantics. A PLSA model was trained by using the expectation maximization algorithm (Dempster, Laird & Rubin, 1977) where convergence of the likelihood is used as a stopping criterion.

**CRBM:** the use of binomial units in CRBM requires tuning the number of units in the hidden layer. We conducted a number of experiments with different latent space dimensions and observed that the performance was insensitive to the dimensionality of latent space. This can be explained by the fact that the CRBM is designed to smooth the term-to-document relatedness rather than term-to-term relatedness. In our experiments, we used the same number of hidden units in CRBM as that used in our CE space (c.f. Figure 2). CRBM models were trained with the contrastive divergence algorithm (Hinton, 2002) where we used the recommended learning rate of 0.1 with moment 0.5. Our implementation is based on the MatRBM[4] package.

**Random**: this is a model used to form a baseline without learning. Depending on an evaluation criterion, the model worked by returning a proper number of terms uniformly sampled from a test data set to form the primed list for a given query term.

Once those models were trained, the following methods were used in semantic priming as well as ranking the different terms for a query instance or a query document:

**PCA, LSA, InfoTheo and Skip Gram**: We have investigated two distance metrics in our experiments, i.e., the cosine and the Euclidian distances. As the cosine metric outperformed Euclidian for all models, we used the cosine metric to measure the distance between different terms in the semantic representation space.

**LDA and PLSA**: The information theoretic distance between a pair of terms given a topic distribution was used (c.f. Equation 1).

**CRBM**: The model was tested with 100 trials by using one hot representation of the query term, i.e., a vector with all zeros except one value representing the query term set to one, via activating the unit corresponding to the query term and presenting one hot representation of the document containing the query term as its context. In each trial, the model acted 100 consecutive forward and backward steps. As a result, all the 100 output vectors on the visible units achieved from 100 trials are averaged and the averaged output of visible units was used to measure relatedness; for the visible units of the higher activation values in output, their corresponding terms are treated as having higher relatedness. Note that due to the technical limitation of this model (c.f. Sections 2.3 and 7), it could be evaluated only on training sets in the WCT experiments.

R**andom**: The model ranks the terms randomly with a uniform distribution.

**Our Model**: The terms are ranked based on their Euclidian distances in the concept embedding (CE) space to a query concept. It should be clarified that the prediction learning in our model may lead to a CE space that facilitates the final CE space formation via distance learning (c.f. Section 4). To evaluate the gain of distance learning with our proposed Siamese architecture, we apply the ***CE*** representations achieved via the initial prediction learning, named *CE,* and the final distance learning, dubbed *Siamese-CE,* to semantic priming.

In summary, our comparative studies in applying different semantic learning models in semantic priming are based on exactly the same experimental settings. While all of the aforementioned semantic learning models were evaluated with the extended priming protocol, only LDA, PLSA, CRBM and Random models along with ours were evaluated in the priming protocol described in Section 6.1 since only these models can generate their semantic representations with a term and its local context simultaneously.

---

[4] https://code.google.com/p/matrbm/

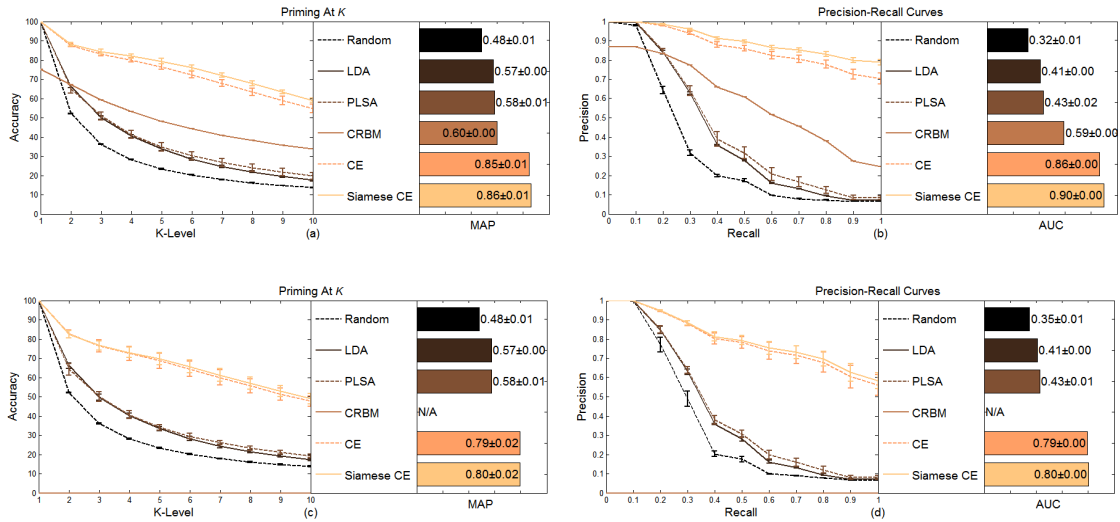

Figure 8: Priming accuracy of different models in terms of **P@K**, MAP, Precision vs. Recall and AUC on MagTag5K. (a-b) Training subset. (c-d) Test subset. Error bars regarding the **P@K** and the precision-recall curves indicate standard error and the numbers regarding the MAP and the AUC are mean and standard error of the MAP and AUC priming accuracy. This notation is applied to all the figures hereinafter.

## 6.3 Within-Corpus Results

With the experimental setting described in Section 6.2.1, we report the WCT experimental results on three training corpora: MagTag5K, CAL500 and Corel5K in terms of two priming protocols.

### 6.3.1 Priming Results

Figure 8 illustrates the priming results of five different models on MagTag5K in terms of four evaluation criteria defined in Section 6.1.3. Figure 8(a) shows the priming results on the training subset in terms of $P@K$ as $K$ varies from one to 10 and the MAP results, both indicated by the mean and standard error on three trials. It is observed from Figure 8(a) that all the models apart from the CRBM always outperform the Random model regardless of $K$ and the length of evaluated docu ments. The CRBM performs worse at $K = 1$ but much better than the Random for different $K$ values up to 10 in terms of $P@K$ and the MAP. Normally, the ground-truth for $P@1$ corresponds to the query term itself and the stochastic nature of CRBM might be responsible for its failure at $K = 1$. Figure 8(b) shows the priming results on the training subset in terms of the precision-recall performance at 11 standard recall levels and the aggregated AUC. The same as seen in Figure 8(a) is observed. It is evident from Figures 8(a) and 8(b) that our model performs the best among all five models regardless of evaluation criteria. In particular, Siamese-CE leads to the significantly better performance by beating the runner-up, the CRBM, with a big margin, e.g., 26% in MAP and 31% in AUC. Also we observe that Siamese-CE performs slightly better than CE on the training subset. Figures 8(c) and 8(d) show the priming results on the test subset in terms of four performance indexes, respectively. The exactly same as seen on the training subset is observed on the test subset although the performance of all the models on the test subset is degraded in comparison to that on the training subset. While the CRBM is no longer applicable to the test subset, Siamese-CE still wins with a big margin of at least 22% in MAP and as least 37% in AUC in comparison to other three models. It is also observed that **CE** representation seems to have a better generalization capability than Siamese-CE although Siamese-CE still performs better than CE on the test subset. Overall, our model outperforms others with the statistical significance (p-value < .01, Student's t-test) apart from $K = 1$. The experimental results on this data set demonstrates that the accurate concepts and their relatedness have been captured by using both terms and their local context and such learnt semantics can be well generalized to those documents that were never seen in training.

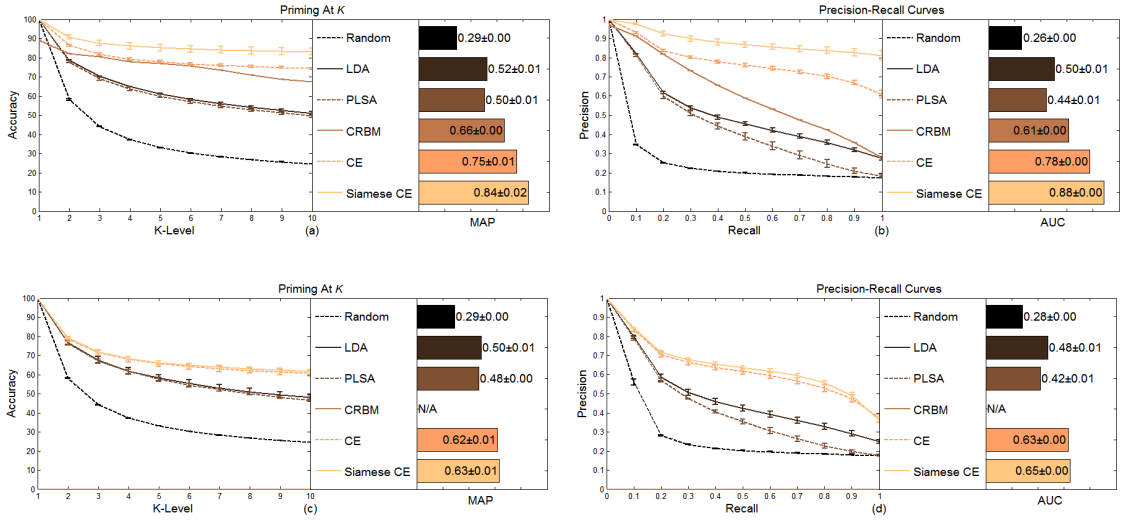

Figure 9: Priming accuracy of different models in terms of **P@K**, MAP, Precision vs. Recall and AUC on CAL500. (a-b) Training subset. (c-d) Test subset.

Figure 9 shows the priming results of five different models on CAL500 in terms of four evaluation criteria. It is observed from Figures 9(a) and 9(b) that our model performs significantly better than other models on the training subset given the fact that Siamese-CE yields at least 18% in MAP and at least 27% in AUC higher accuracy than other models. It is also observed that the high document cardinality of this data set makes the Random model relatively easy to guess a few related terms, i.e., results in relatively high $P@K$ for small $K$ values, as evident in Figure 9(a). In Figure 9(b), it is seen that higher precision at high recall levels is achieved than that achieved on the training subset in MagTag5K. As a runner-up, however, the performance of CRBM decreases rapidly as the recall level increases. This suggests that the CRBM had encountered a difficulty in identifying all the terms related to a query concept. The same problem can be found in other models except ours. Figures 9(c) and 9(d) illustrate the performance of different models on the test subset. Overall, the same conclusions drawn on the training subset are reached on the test subset; Siamese-CE yields the statistically significant better performance (p-value < .01, Student's t-test) than other models by winning at least 13% in MAP and at least 17% in AUC on the test subset. In comparison to the results on MagTag5K shown in Figure 8, our model generally behaves consistently though the generalization performance on CAL500 is worse than that on MagTag5K. As described in Section 5.1, CAL500 is a music tag collection quite different from MagTag5K in terms of length of documents or document cardinality and the tag usage distribution (c.f. Figure 3). In light of capturing the accurate concepts and their relatedness, the experimental results on two distinct music data sets suggest that our model is not sensitive to document cardinality and statistics underlying different collections in the same domain as is evident in Figures 8 and 9.

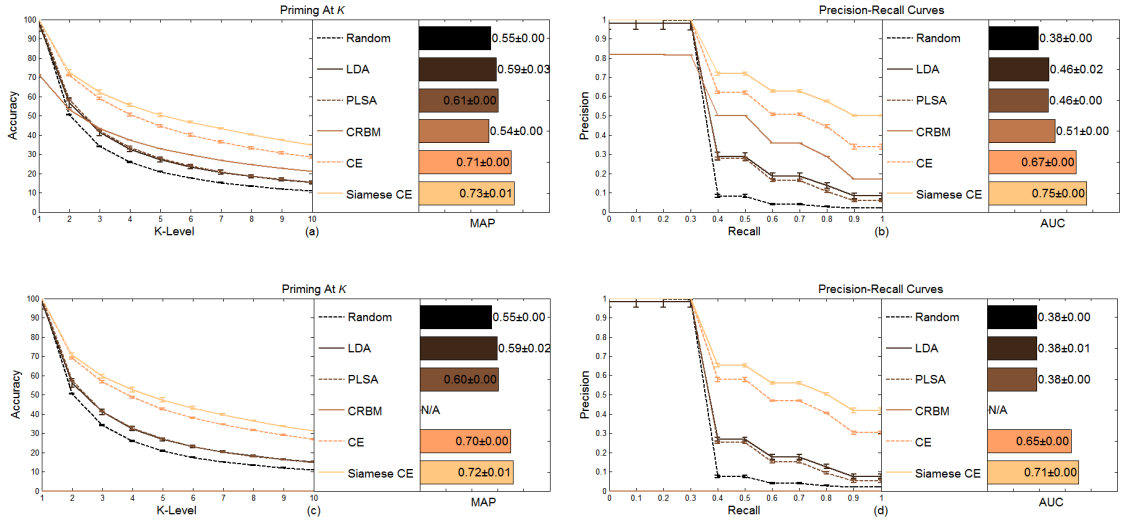

Figure 10: Priming accuracy of different models in terms of ***P@K***, MAP, Precision vs. Recall and AUC on Corel5K. (a-b) Training subset. (c-d) Test subset.

Figure 10 illustrates the priming results of five different models on Corel5K in the image domain in terms of four evaluation criteria. Overall, our model yields the statistically significant better results (p-value < .01, Student's t-test) than other models. On the training subset, it is evident from Figures 10(a) and 10(b) that Siamese-CE leads to at least 12% in MAP and at least 24% in AUC higher than others. From Figures 10(c) and 10(d), the favorable generalization capability of our model is seen clearly; on the test subset, Siamese-CE considerably outperforms other models by winning at least 12% in MAP and 33% in AUC. In addition, the gain of Siamese-CE over CE is more visible on this data set. From Figure 10, however, it is also observed that the performance is degrading rapidly across the ranked list due to the nature of this data set. As described in Section 5.1, a document in this data set contains only five labels at maximum and 3.5 labels on average, but there is a vocabulary of 292 different labels in this data set. Once the value of $K$ in $P@K$ and the recall level reach a certain degree beyond the length of a query document, the performance is inevitably degraded regardless of which model is used. Even in this situation, the experimental results shown in Figure 10 suggest that our model still yields the significantly better performance, in particular, at high recall levels, as is evident in Figure 10(b) and Figure 10(d). In general, the experimental results on this data set demonstrate the capability of our model in capturing the accurate concepts and their relatedness from documents containing only a small number of terms.

### 6.3.2 Extended Priming Results

Figure 11 illustrates the extended priming results of nine different models on MagTag5K in terms of four evaluation criteria defined in Section 6.1.3. Regarding the results on the training subset shown in Figures 11(a) and 11(b), our model always outperforms all other models with the statistical significance (p-value < .01, Student's t-test) in all four evaluation criteria. In particular, Siamese-CE wins at least 26% in MAP and 34% in AUC over other models. As shown in Figures 11(c) and 11(d), our model also performs the best on the test subset, and moreover, Siamese-CE beats the runner-up with a big margin of 23% in MAP and 22% in AUC. On both training and test subsets, our model performs particularly well at high recall intervals as shown in Figures 11(b) and 11(d). Overall, CE performs equally well on both training and test subsets in MagTag5K. A closer look suggests that Siamese-CE outperforms CE at high recall levels on the training set but this advantage disappears on the test subset. The better performance achieved by Siamese-CE on the training subset is thanks to the distance learning that refines the ***CE*** representation. On both training and test subsets, LSA, Skip Gram, LDA and PLSA all perform poorly although they win over the Random model. Interestingly, LDA and PLSA are two probabilistic topic models (PTMs) that yield document-level representations. In this document-level priming evaluation, however, the PTMs do not seem to be able to capture the subtle difference in the concepts conveyed in a query document, which provides evidence to support our contextualized semantic learning problem formulation. From Figure 11, it is also evident that Skip Gram cannot capture the semantics from tags well due to a lack of syntactic context in documents of descriptive terms. In contrast, the non-contextualized models, PCA and InfoTheo, perform well given the fact they win over almost all other models apart from ours as illustrated in Figure 11. On the training subset, however, CRBM performs better than PCA and InfoTheo at both small $K$ in *P@K* and low recall intervals, as shown in Figures 11(a) and 11(b), due to its capability in capturing document-term relatedness. It is worth stating that the success of PCA and InfoTheo relies on the careful weighting of the document-term matrix and proper aggregation and those results reported here are those corresponding to the optimal parameters. Finally, the experimental results on MagTag5K in both priming and extended priming shown in Figures 8 and 11 also raise an issue on why the distance learning by our Siamese architecture does not lead to a substantial gain on this data set, in particular, regarding generalization, which will be discussed later on.

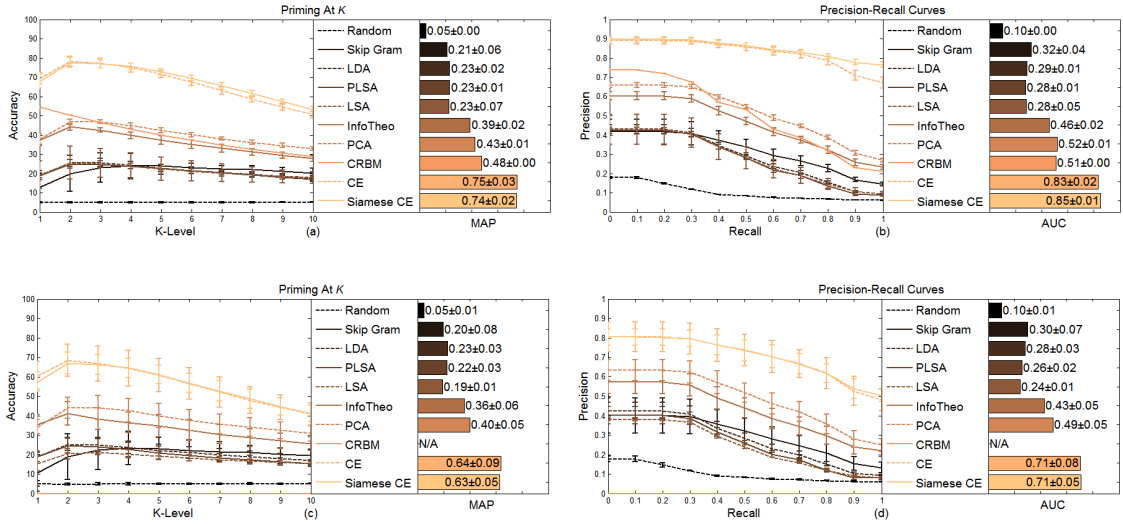

Figure 11: Extended priming accuracy of different models in terms of ***P@K***, MAP, Precision vs. Recall and AUC on MagTag5K. (a-b) Training subset. (c-d) Test subset.

Figure 12 shows the extended priming results of nine different models on CAL500 in terms of four evaluation criteria. Once again, our model outperforms other models regardless of evaluation criteria. As shown in Figures 12(a) and 12(b), the results on the training subset indicate that Siamese-CE wins over other models at least 18% in MAP and at least 25% in AUC and, in particular, our model performs much better at high recall levels. In comparison to results on MagTag5K, there are two non-trivial changes: CRBM outperforms PCA and InfoTheo considerably and Siamese-CE performs significantly better than CE on the training subset of this data set. Nevertheless, the results on the test subset shown in Figures 12(c) and 12(d) reveal that all the models including ours seem to face difficulty in extended priming especially at high recall levels. The difficulty causes the performance of some models, e.g., LDA and PLSA, to be close to that of the Random model. An analysis on the training subset reveals that it may be caused by a lack of sufficient informative training examples given the fact that 335 training documents actually consist of 158 different tags. Due to insufficient training data reflecting various concepts and intended terms' use patterns, it is likely that the learning may overfit the training data and hence some unseen positive instances may be grouped incorrectly with negative instances in our distance learning. As a consequence, our winning margin over other models becomes smaller in comparison to results on MagTag5K (c.f. Figures 11(a) and 11(b)), e.g., Siamese-CE gains only 5% higher in MAP and nothing in AUC in comparison to the runner-up, PCA. As a non-contextualized model, PCA learns the global relatedness of tags. In the presence of insufficient training documents for capturing the accurate concepts, the PCA may be a choice after trade-off between the performance gain and the computational efficiency in this document-level retrieval task.

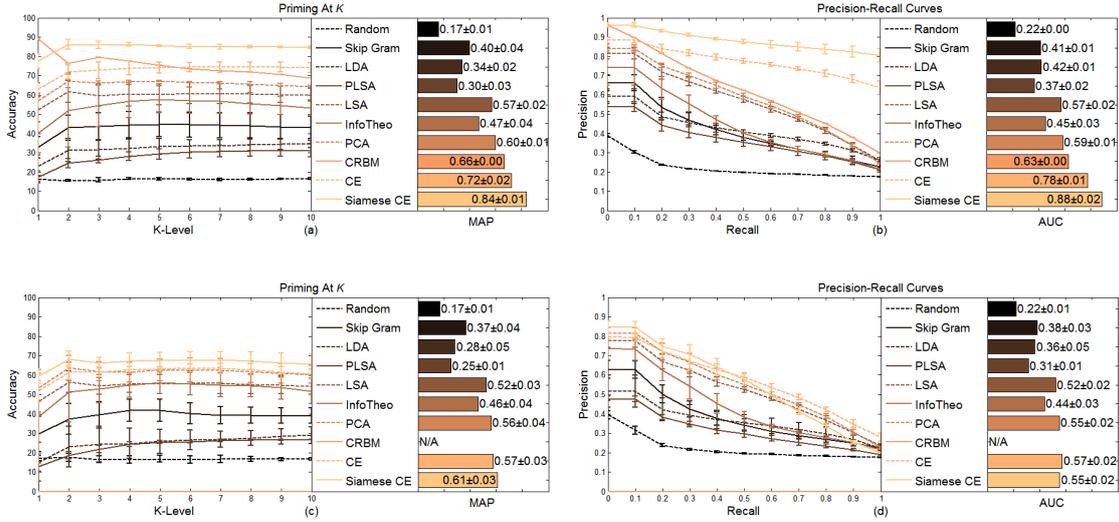

Figure 12: Extended priming accuracy of different models in terms of ***P@K***, MAP, Precision-Recall and AUC on CAL500. (a-b) Training subset. (c-d) Test subset.

Figure 13 illustrates the extended priming results of nine different models on Corel5K in the image domain in terms of four evaluation criteria. Overall, our model yields the statistically significant better performance e (p-value < .01, Student's t-test) than other models on both training and test subsets. As shown in Figures 13(a) and 13(b), the results suggest that Siamese-CE wins over other models at least 10% in MAP and at least 20% in AUC on the training subset. Once again, non-contextualized models, PCA and InfoTheo, outperform other models apart from ours. Figures 13(c) and 13(d) illustrate the results on the test subset where all the models are ranked as same as done on the training subset in in terms of their performance. Siamese-CE wins over the runner-up, InfoTheo, 9% in MAP and 15% in AUC and also leads to better generalization than CE with the gain of 5% in MAP and 9% in AUC. In particular, Siamese-CE outperforms CE at high recall levels in both training and test subsets as shown in Figures 13(a) and 13(b). It indicates that the distance learning would be paid off should there be sufficient informative training examples regarding various concepts and intended term-use patterns. Once again, LDA and PLSA perform poorly on this data set as is evident in Figure 13, which lends further evidence to support our contextualized semantic learning given the fact that a huge gain is brought by our model based on LDA.

In summary, the WCT experimental results on different data sets in two different priming protocols demonstrate that our approach generally outperforms other state-of-the-art methods in semantic priming and has the proven generalization capability that the learnt semantics can be applied to unseen documents in training for this retrieval task.

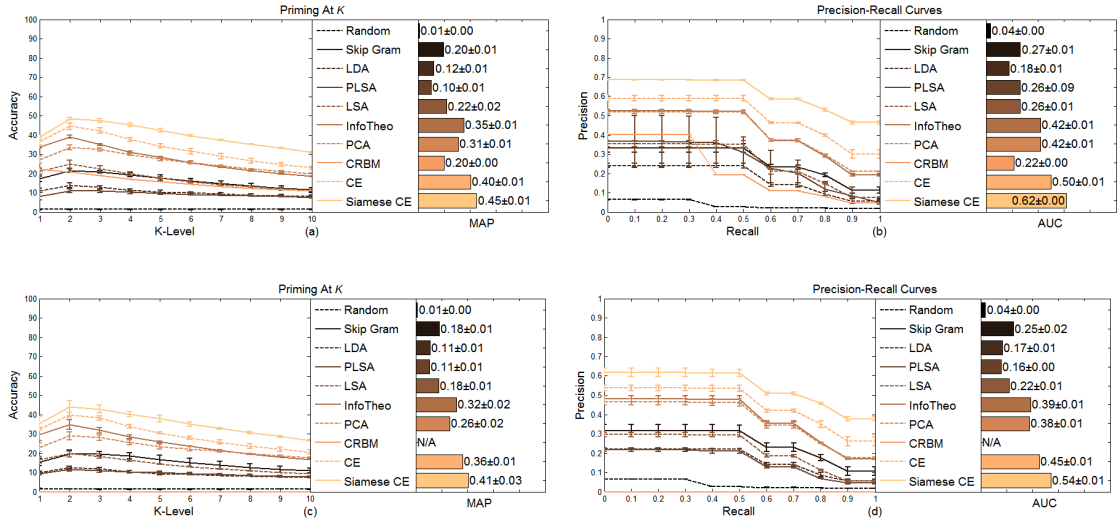

Figure 13: Extended priming accuracy of different models in terms of **P@K**, MAP, Precision vs. Recall and AUC on Corel5K. (a-b) Training subset. (c-d) Test subset.

### 6.4 Cross-Corpora Results

In the CCT experiments, we apply the semantic representation achieved by a model trained on a corpus to another test collection for semantic priming. Here, we report results for those semantics trained on MagTag5K and applied to MSD as well as those trained on Corel5K and applied to LabelMe and SUNDatabase in terms of two priming protocols.

#### 6.4.1 Priming Results

Figure 14 illustrates the priming results of four different models on three test collections in terms of four evaluation criteria defined in Section 6.1.3.

Figures 14(a) and 14(b) show the priming results on MSD. Overall, our model outperforms other model with the statistical significance (p-value < .01, Student's t-test). It is observed from Figures 14(a) and 14(b) that Siamese-CE gains at least 9% in MAP and at least 19% in AUC higher than other models and, in particular, yields the considerably better performance at high recall levels. Also CE leads to a considerably better performance than other models and its performance is slightly lower than that of Siamese-CE. In contrast, LDA and PLSA yield the results close to those generated by the Random model, which indicates the poor transferability of semantics learnt by two models. By comparison to the results on the test subset of MagTag5K shown in Figures 8(c) and 8(d), we observe that the performance of Siamese-CE on MSD is worse than the WCT results, e.g., 12% in MAP and 14% in AUC lower. As MagTag5K is a small subset of MSD, it is likely that there are much more varied patterns and concepts associated with a tag in MSD and different annotators working on the large collection, which could have more and alternative interpretations or intentions for those tags in MagTag5K in a much larger tag vocabulary in MSD. Despite the lower performance on MSD, we believe that the priming results generated by our model are quite reasonable and promising in learning transferable semantics.

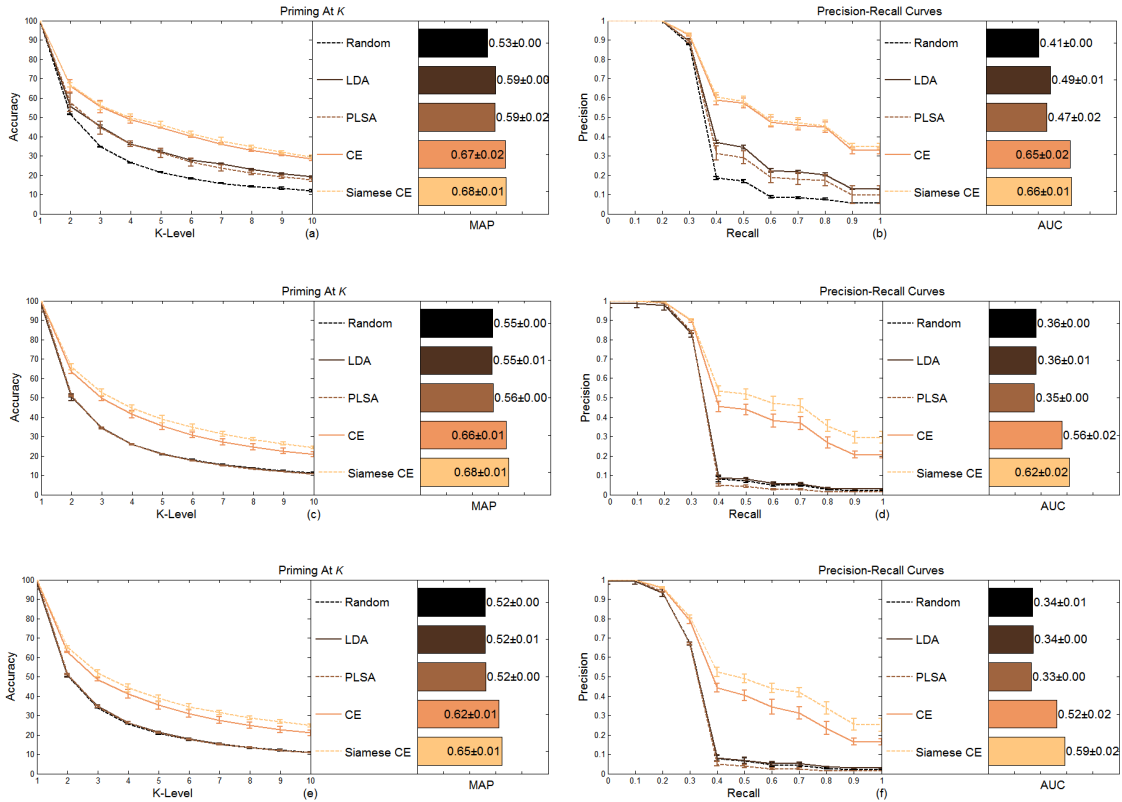

Figure 14: Priming accuracy of different models in terms of ***P@K***, MAP, Precision vs. Recall and AUC in the CCT experiments. (a-b) MSD. (c-d) LabelMe. (e-f) SUNDatabase.

Figures 14(c) and 14(d) show the priming results on LabelMe. It is observed that our model outperforms other model with the statistical significance (p-value < .01, Student's t-test); Siamese-CE wins over the runner-up 12% in MAP and 27% in AUC and, in particular, the significantly better performance at high recall levels. Also the performance of CE is superior to that of other models but lower than that of Siamese-CE. Unfortunately, LDA and PLSA yield rather poor performance, roughly identical to that of the Random model, as clearly seen in Figures 14(c) and 14(d). In contrast to the results on the test subset of Corel5K shown in Figures 10(c) and 10(d), it is observed that that the performance of Siamese-CE on LabelMe is close to the WCT results, e.g., only 4% in MAP and 9% in AUC lower. Also it still maintains the good performance at high recall levels. Those results suggest quite strongly that our model can capture the transferable semantics when training and test corpora have a high agreement in intended meanings of terms in annotation.

Figures 14(e) and 14(f) show the priming results on SUNDatabase. Once again, our model outperforms other model with the statistical significance (p-value < .01, Student's t-test); Siamese-CE wins over the runner-up 13% in MAP and 25% in AUC and, in particular, the significantly better performance at high recall levels. It is also observed that all the models perform on this data set very similarly to those on LableMe although Siamese-CE yields a lower priming accuracy in comparison to that on LabelMe, e.g., 3% in both MAP and AUC.

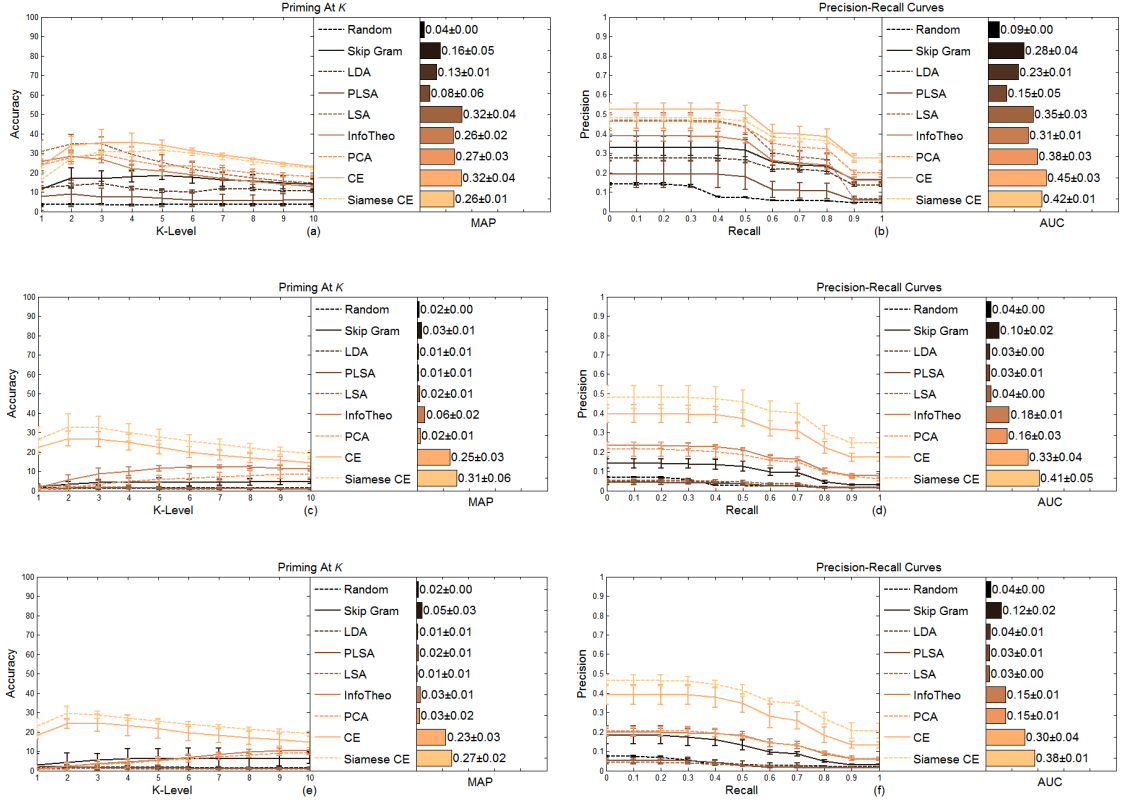

Figure 15: Extended priming accuracy of different models in terms of ***P@K***, MAP, Precision vs. Recall and AUC in the CCT experiments. (a-b) MSD. (c-d) LabelMe. (e-f) SUNDatabase.

For three image data sets, we notice that the document cardinality is quite different given the fact that on average there are 3.5, 7.3 and 11 labels per document in Corel5K, LabelMe and SUNDatabase, respectively. This information implies that our model is less sensitive to some statistical variation but more sensitive to the semantics underlying co-occurring terms.

### 6.4.2 Extended Priming Results

Figure 15 shows the extended priming results of nine different models on three test collections in terms of four evaluation criteria defined in Section 6.1.3.

Figures 15(a) and 15(b) illustrate the extended priming results on MSD. In comparison to the performance on the test subset on MagTag5K shown in Figures 11(c) and 11(d), all the models including ours perform poorly, which demonstrates the challenge in learning transferrable semantics with limited training data. It is observed that LSA performs the best in MAP while our model wins in AUC. In general, our model performs better at large *K* and high recall levels while LSA outperforms others at small *K* and low recall levels. In particular, CE always outperforms Siamese-CE. For the reason described in Section 6.4.1, a contextualized model is more sensitive to the usage patterns and intended meanings of terms in capturing concepts in context than a non-contextualized model that learns only global relatedness. In general, both the priming and the extended priming results on MSD suggest that a contextualized semantic model does not seem to transfer the semantics learnt from a less informative data set to those of richer information, intricate concepts and alternative intended term-use patterns.

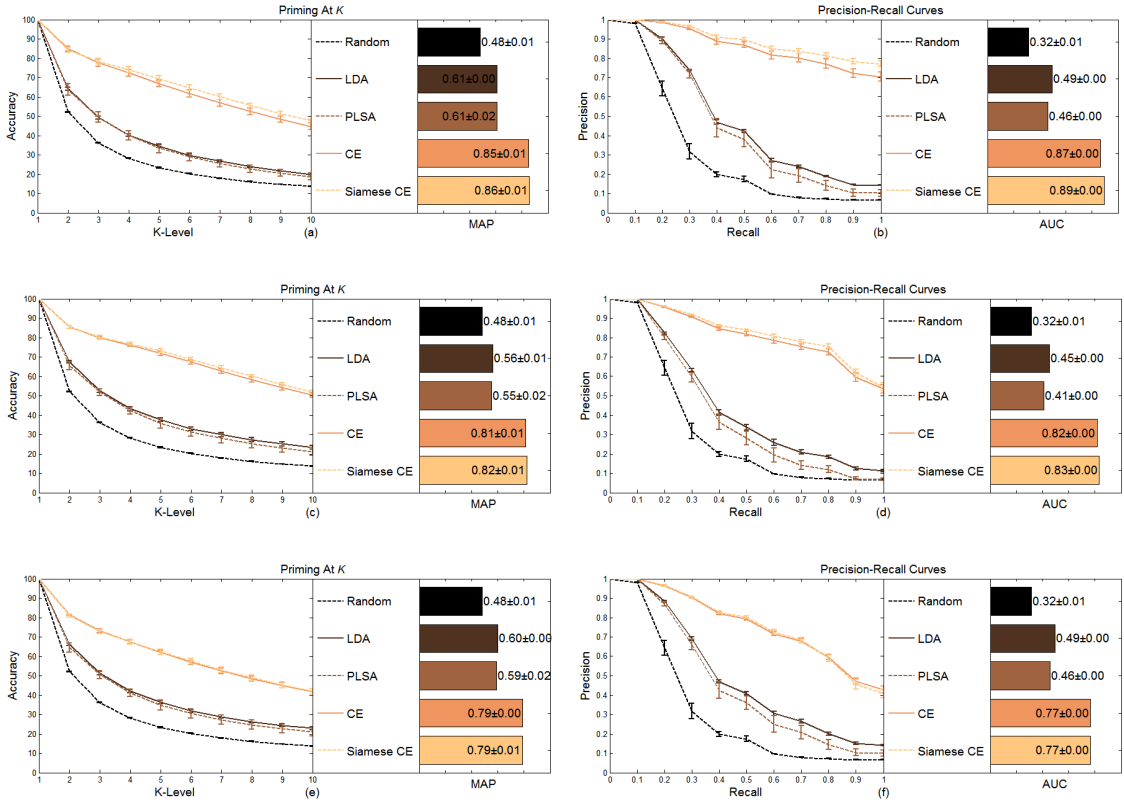

Figure 16: Priming accuracy of different models on MagTag5K at three missing rates. (a-b) Up to 10%. (c-d) Between 10% and 30%. (e-f) Between 30% and 50%.

Figures 15(c) and 15(d) show the extended priming results on LabelMe. It is observed that our model outperforms other models with the statistical significance (p-value < .01, Student's t-test). In general, the behavior of our model on this data set is remarkably similar to that on the test subset of Corel5K as shown in Figures 13(c) and 13(d) and Siamese-CE always performs better than CE. Unfortunately, all other models generally perform poorly; most of models yield the performance roughly identical to that of the Random model, as clearly seen in Figures 15(c) and 15(d). In general, the performance of our model is consistent in both the priming and the extended priming on this data set. Hence, the same conclusion on the priming can be drawn on the extended priming.

Figures 15(e) and 15(f) show the extended priming results on SUNDatabase. Once again, our model performs statistically significant (p-value < .01, Student's t-test) better than all other models; Siamese-CE wins over the runner-up 22% in MAP and 13% in AUC and, in particular, the significantly better performance at all 11 recall levels. It is also observed that all the models perform on this data set very similarly to those on LableMe.

In summary, the CCT experimental results demonstrate that the semantics learnt by our model trained on a data set may be transferable to other collections if different annotators have a high agreement on the intended meanings of terms and there are sufficient training documents reflecting various concepts and intended term-use patterns. Without meeting the requirement, all the models encounter the same problem in generalization of leant semantics across corpora.

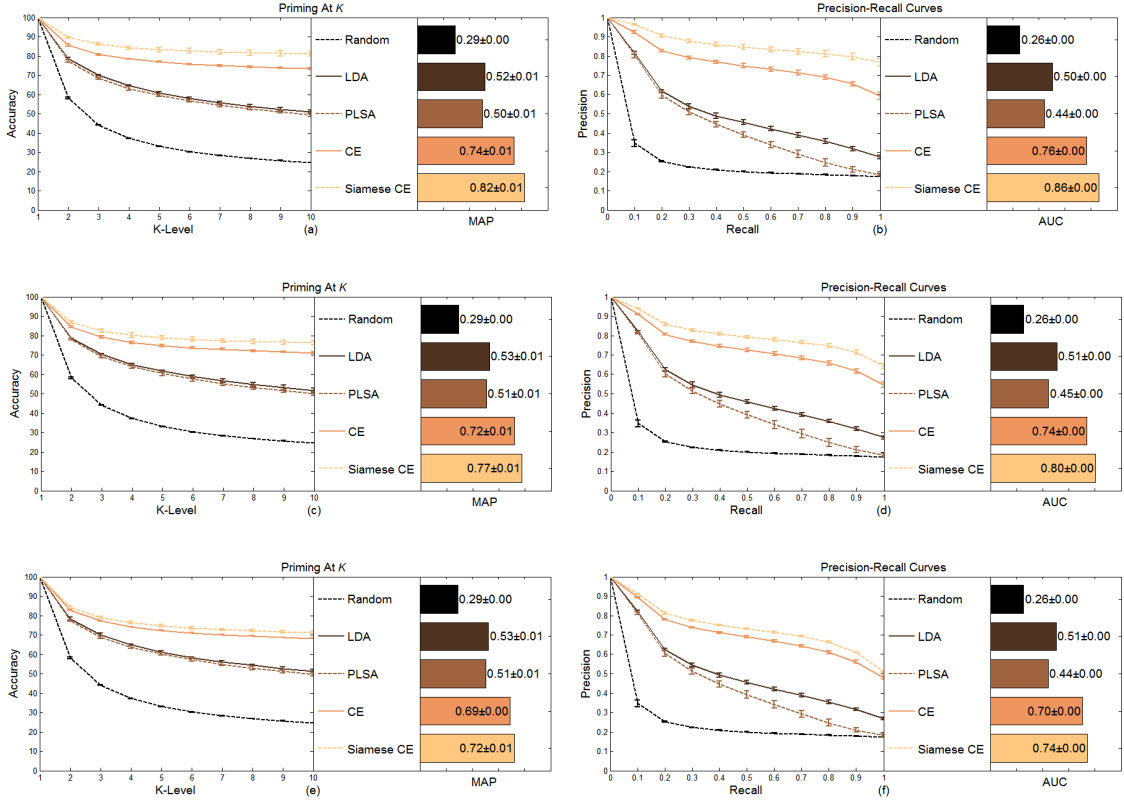

Figure 17: Priming accuracy of different models on CAL500 at three missing rates. (a-b) Up to 10%. (c-d) Between 10% and 30%. (e-f) Between 30% and 50%.

### 6.5 Incomplete Local Context Results

In the incomplete local context experiments, we randomly remove a number of terms from an evaluation document to synthesize an incomplete local context with three missing rates, up to 10%, between 10% and 30% and between 30% and 50%, as discussed in section 6.2.3. The training subsets in MagTag5K, CAL500 and Corel5K are used in this experimental setting and we report the experimental results in terms of the priming and the extended priming protocols. It is also worth clarifying that the CRBM is generally ineligible as its local context is the ID of a query document and hence cannot be distorted. Nevertheless, we use the CRBM only in the extended priming protocol although its local context is not distorted.

#### 6.5.1 Priming Results

Figure 16 illustrates the priming results of four different models at three missing rates on MagTag5K in terms of four evaluation criteria defined in Section 6.1.3. As expected, it is observed from Figure 16 that the use of incomplete local context results in the degraded performance for our model due to information loss. In comparison to the results with the complete local context, the performance of Siamese-CE shown in Figure 16 is lower than those shown in Figure 8(a) and 8(b) by 0%, 4% and 7% in MAP as well as 1%, 7% and 13% in AUC at three missing rates, respectively. In particular, the incomplete local context generally causes the performance at high recall levels to be degraded more than that at low recall levels. In contrast, two PTMs, LDA and PLSA, show their robust performance in resisting noisy data as their performance on the incomplete documents is even better than that on the corresponding complete version. Nevertheless, our model still outperforms other models with the statistical significance (p-value < .01, Student's t-test) as clearly seen in Figure 16.

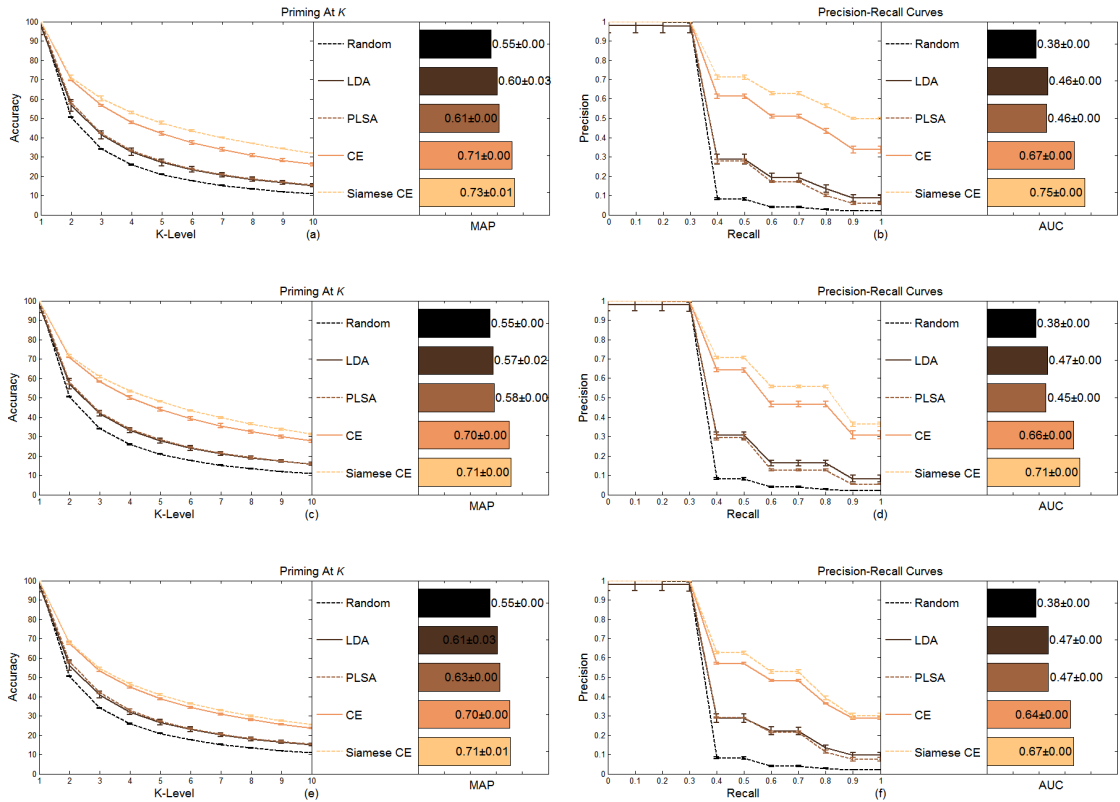

Figure 18: Priming accuracy of different models on Corel5K at three missing rates. (a-b) Up to 10%. (c-d) Between 10% and 30%. (e-f) Between 30% and 50%.

Figure 17 shows the priming results of four different models at three missing rates on CAL500 in terms of four evaluation criteria. On this data set, all the models exhibit almost the same behavior as they work on MagTag5K in the presence of incomplete local context. In comparison to the results with the complete local context, the performance of Siamese-CE on this data set is reduced by 2%, 7% and 12% in MAP as well as 2%, 8% and 14% in AUC at three missing rates, respectively. Unlike the behavior on MagTag5K, however, it is observed from Figure 17 that the performance of Siamese-CE at high recall levels does not decrease sharply. Although LDA and PLSA yield the robust performance, our model still generates the statistically significant (p-value < .01, Student's t-test) better performance than all other models including LDA and PLSA in this experimental setting, as is evident in Figure 17.

Figure 18 shows the priming results of four different models at three missing rates on Corel5K in terms of four evaluation criteria. It is observed from Figure 18 that our model exhibits the better robustness in the presence of incomplete local context. On this data set, the performance of Siamese-CE is only reduced by 0%, 2% and 2% in MAP as well as 0%, 4% and 8% in AUC at three missing rates, respectively, in comparison to those with complete local context shown in Figures 10(a) and 10(b). Recall that on average there are only 3.5 labels in this data set. By dropping up to 50% terms per document, on average, there are less than two labels per document to form local context. Thus, our model leads to favorable results on this data set. Once again, our model yields the statistically significant (p-value < .01, Student's t-test) better performance than other models on Corel5K, as is evident in Figure 18.

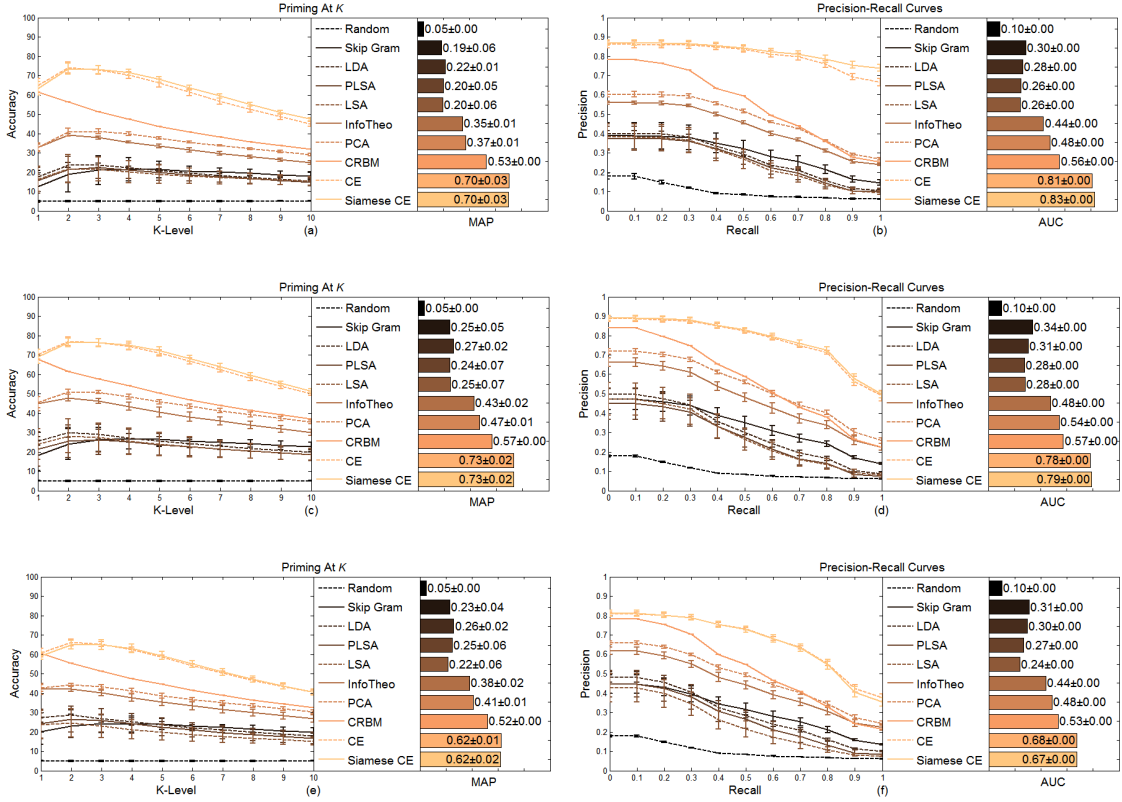

Figure 19: Extended priming accuracy of different models on MagTag5K at three missing rates. (a-b) Up to 10%. (c-d) Between 10% and 30%. (e-f) Between 30% and 50%.

### 6.5.2 Extended Priming Results

Figure 19 illustrates the extended priming results of nine different models at three missing rates on MagTag5K in terms of four evaluation criteria defined in Section 6.1.3. It is observed from Figures 16 and 19 that the behavior of our model generally remains consistent in two different priming protocols; the performance gradually decreases as the missing rate increases and a higher missing rate causes Siamese-CE to have a sharper performance reduction at high recall levels. With the incomplete local context, the performance of Siamese-CE is reduced by 4%, 1% and 12% in MAP as well as 2%, 6% and 18% in AUC at three missing rates, respectively, in comparison to those with complete local context. It is also observed from Figure 19 that unlike our model, other models perform irregularly, e.g., the performance at a higher missing rate is even better than that at a lower missing rate. Overall, their performance is still significantly inferior to ours.

Figure 20 shows the extended priming results of nine different models at three missing rates on CAL500 in terms of four evaluation criteria. We observe that on this data set, all the models including ours behave similarly in comparison to their behavior on MagTag5K; all other models behave irregularly at different missing rates while the performance of our model gradually decreases as the missing rate increases. With the incomplete local context, the performance of Siamese-CE is reduced by 1%, 4% and 10% in MAP as well as 2%, 5% and 16% in AUC at three missing rates, respectively, in comparison to those with the complete local context. Thanks to the use of the complete local context, i.e., the document ID, CRBM performs well at all three missing rates although it is still inferior to ours overall. Despite the performance reduction, our model still outperforms all other models on this data set and, in particular, can prime a few top related terms correctly, as is evident in Figure 20.

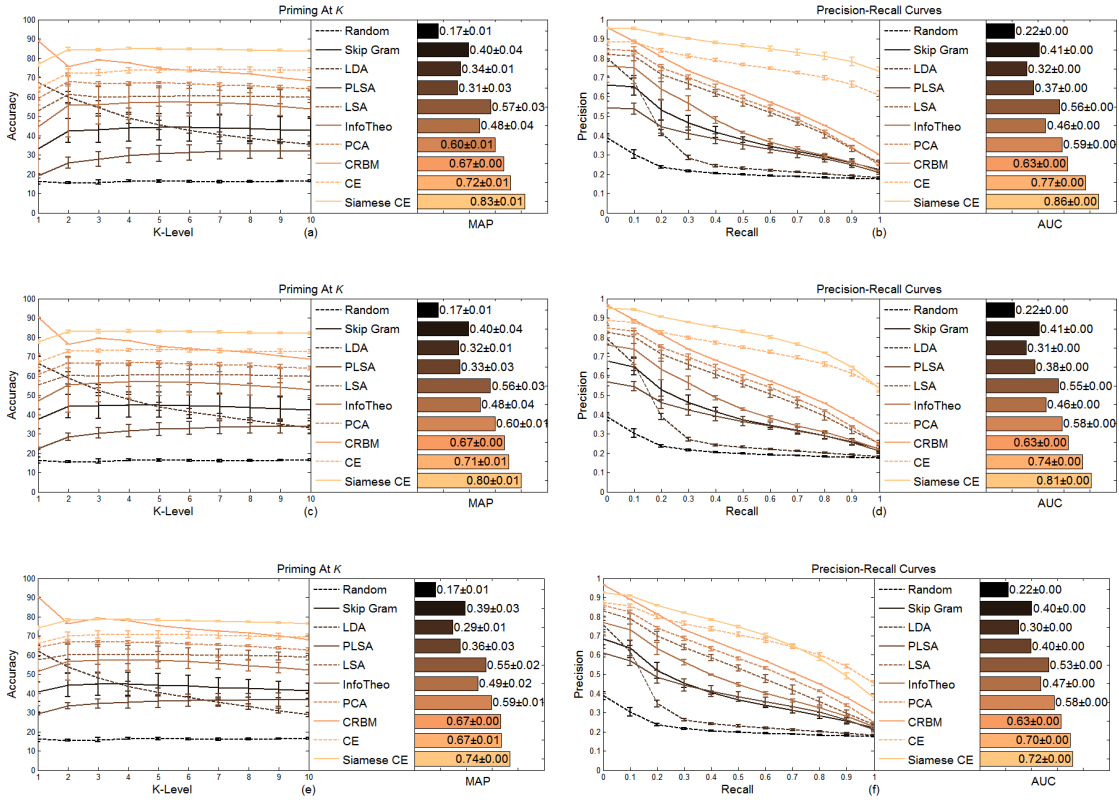

Figure 20: Extended priming accuracy of different models on CAL500 at three missing rates. (a-b) Up to 10%. (c-d) Between 10% and 30%. (e-f) Between 30% and 50%.

Figure 21 illustrates the extended priming results of nine different models at three missing rates on Corel5K in terms of four evaluation criteria. Unlike its behavior on two music data sets, our model performs much better at two low missing rates and reasonably at the highest missing rate; the performance of Siamese-CE is only reduced by 1%, 1% and 10% in MAP as well as 1%, 4% and 14% in AUC at three missing rates, respectively, in comparison to those with the complete local context. On this data set, our model also behaves consistently in two different priming protocols as shown in Figures 18 and 21. In contrast, all other models have the same behavior as they exhibit on two music data sets. In general, our model still outperforms other models at all three missing rates in terms of all four evaluation criteria.

In summary, the experimental results in the presence of incomplete local context suggest that our model performs reasonably well despite the performance reduction as expected and generally outperforms all other models on three different data sets in both the priming and the extended priming protocols. In contrast, other models perform irregularly at different missing rates. It is well known that accompany terms in a document may not convey equal amount information. Dropping a term randomly may exert different impacts on its local context. On the one hand, it could incur a huge information loss and even concept change if the term is very informative. On the other hand, it could make little impact if the removed term is redundant or less informative. In general, removing terms randomly from a document may even cause the loss in coherence of co-occurring terms in an incomplete document. Perhaps this setting might be responsible for irregular yet unstable behaviour of other models and ours on a number of occasions, e.g., on CAL500.

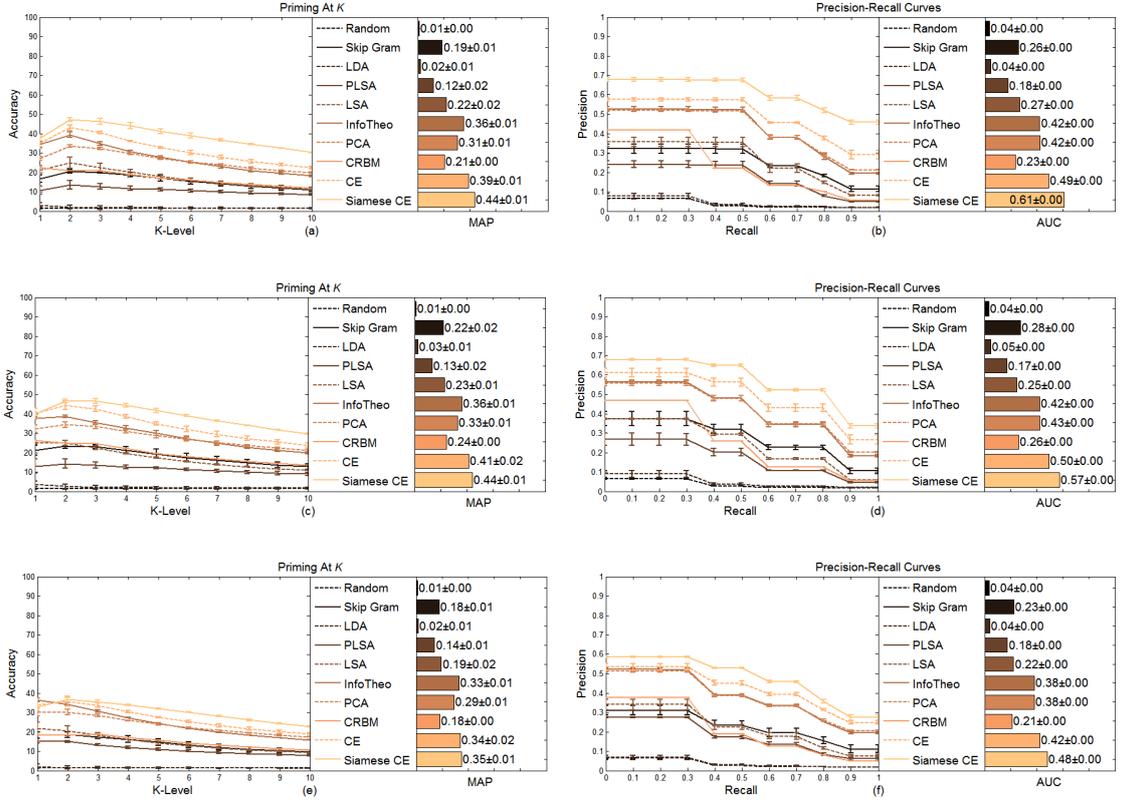

Figure 21: Extended priming accuracy of different models on Corel5K at three missing rates. (a-b) Up to 10%. (c-d) Between 10% and 30%. (e-f) Between 30% and 50%.

## 6.6 Out of Vocabulary Results

With the experimental setting described in Section 6.2.4, we report the OOV experimental results on the reserved OOV set in MagTag5K and the OOV sets consisting of real documents containing OOV terms in MSD, LabelMe and SUNDatabase in terms of two priming protocols. We use $CE(\boldsymbol{\tau_{oov}})$ and Siamese-$CE(\boldsymbol{\tau_{oov}})$ to indicate the representations achieved by the feature-based OOV method, i.e., priming related terms using query concept projection $CE(\tau_{oov}|\delta_{iv})$, and $CE(\boldsymbol{Avg})$ and Siamese-$CE(\boldsymbol{Avg})$ to denote the representations achieved by the concept-based OOV method, i.e., priming related terms using query concept projection, $CE\big(x(\tau_{oov},\delta_{iv})\big)$ (c.f. Section 4.4).

Figure 22 illustrates the priming results of two OOV methods on four different data sets in terms of four evaluation criteria defined in Section 6.1.3. It is evident from Figure 22 that two proposed OOV methods yield favorable results on different data sets, and $CE(\boldsymbol{Avg})$ and

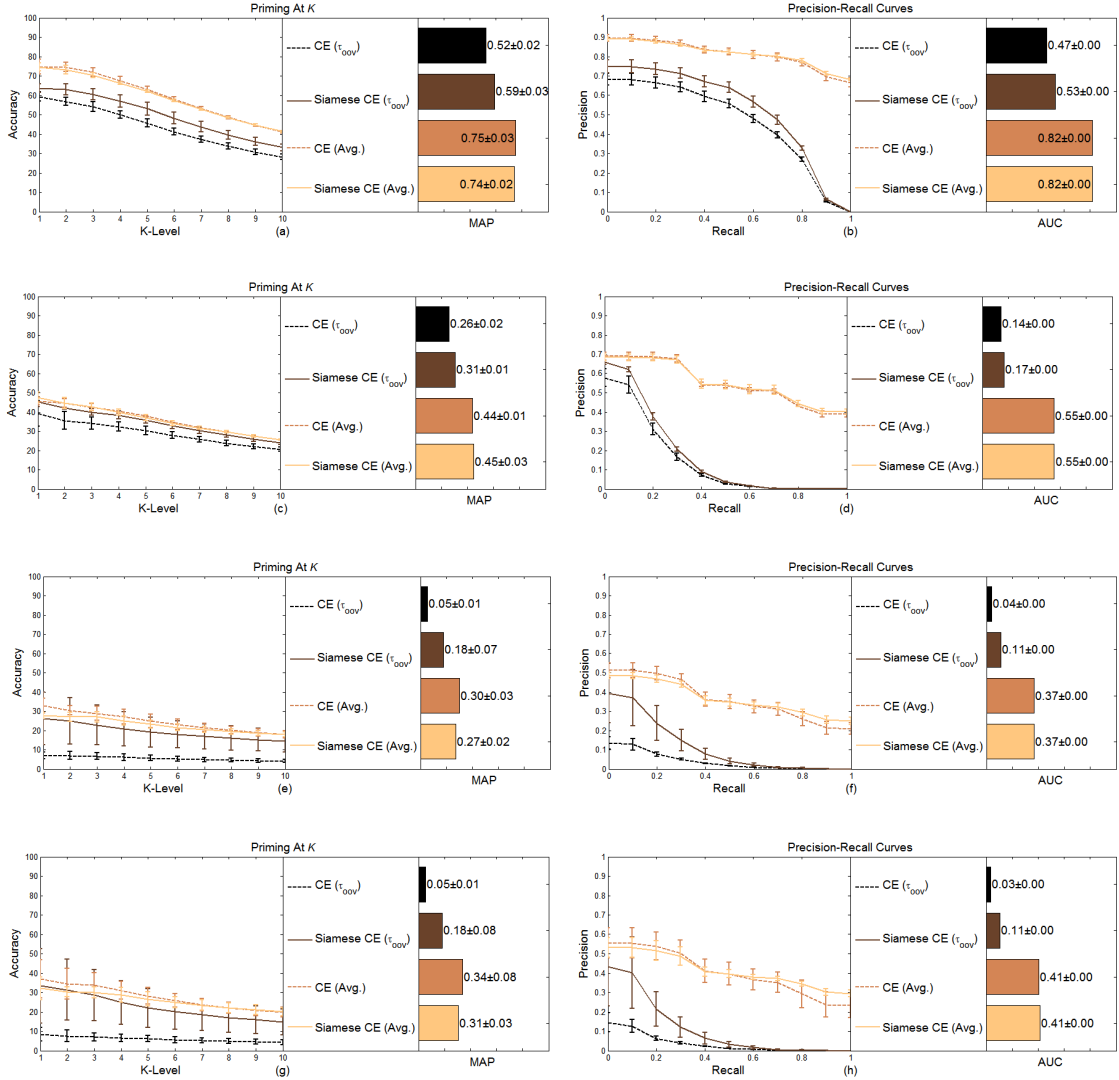

Figure 22: Priming accuracy of two OOV methods on different datasets. (a-b) MagTag5K. (c-d) MSD. (e-f) LabelMe. (g-h) SUNDatabase.

Siamese-CE($Avg$) significantly outperform CE($\tau_{oov}$) and Siamese-CE($\tau_{oov}$) constantly on all four data sets. In particular, the performance of CE($Avg$) and Siamese-CE($Avg$) on the OOV set is roughly comparable to the performance of CE and Siamese-CE on the test subset of MagTag5K, as is evident in Figures 8(c), 8(d), 22(a) and 22(b). For the feature-based method, Siamese-CE($\tau_{oov}$) performs considerably better than CE($\tau_{oov}$) on all four data sets thanks to the distance learning undertaken by our Siamese architecture. However, the performance of CE($Avg$) is marginally higher than that of Siamese-CE($Avg$) on all four data sets. It implies that a subnetwork trained for the prediction has embedded the related concepts conveyed in a document reasonably well so that their centroid in the CE space can be used to approximately embed a related OOV term that shares the same local context. On the other hand, the distance learning undertaken by the Siamese architecture is dedicated to the accurate concept embedding concerning only in-vocabulary terms based on training data. Based on the experimental results shown in Figure 22, we could make better use of CE and Siamese-CE; we use CE to represent

OOV terms while Siamese-CE is used only for in-vocabulary terms. Here, we emphasize that our proposed OOV methods directly generate the same *CE* representation for an OOV term as that of an in-vocabulary term. For any applications that employ our learnt semantics, there is no additional processing required for OOV terms. In other words, both in-vocabulary and OOV terms can be represented uniformly in the CE space.

In summary, all the experimental results reported in Sections 6.3-6.6 provide the solid evidence to support our problem forumulation and the proposed solution as our approach significantly outperforms other state-of-the-art semantic learning methods in this thorough evaluation. In semantic priming, our approach exhibits its strength in capturing accurate semantics from training corpora and, more importantly, the capability of generalizing the learnt semantics to unseen documents in different situations, noisy documents (resulting in incomplete local context) and documents containing OOV terms. Thus, we believe that our approach is ready for different MMIR applications.

## 7 Discussion

In this section, we discuss several issues arising from our work and relate our approach to previous work in learning semantics from descriptive terms.

To learn semantics from descriptive terms, most of existing techniques often undergo a preprocessing stage by filtering out rarely used terms from those documents concerned (Law, Settles & Mitchell, 2010; Mandel et al., 2011). Our observations suggest that some rarely used terms may play a critical role in local context to facilitate understanding the accurate meaning of a specific term. Therefore, our approach always uses all the natural documents without removing any rarely used terms in our training and test. On the other hand, we realize that the frequently used terms may convey commonly used semantics and hence need to be handled differently from rarely used terms. Technically, this could be done in a semantic representation space by using the frequency of terms to normalize the distance between different terms. As this is an issue relating to a specific application, this potential solution needs to be investigated when our approach is applied to a specific task.

In our approach, a novel Siamese architecture and its two-stage learning procedure are proposed especially for learning the concept embedding (CE) from co-occurring terms. As a result, two *CE* representations, CE and Siamese-CE, could be obtained in the first and the second learning stages, respectively; i.e., CE learnt from the prediction task and Siamese-CE generated by working on the distance learning and the prediction task simultaneously. From the experimental results in semantic priming, we observe that Siamese-CE outperforms CE whenever there are sufficient training data reflecting different concepts conveyed in descriptive terms and various intended term usage patterns, while CE may perform slightly better than Siamese-CE for unseen documents in different corpora if the training data do not comply with the aforementioned conditions, e.g., training on MagTag5K, a small music annotation data set, and test on MSD, a huge and highly diversified music annotation data set. As the distance learning for Siamese-CE is dedicated to accurate concept embedding based on information carried in training documents, the lack of sufficient information on various concepts and intended usage patterns in training data is responsible for this problem. In contrast, CE does not involve in the refinement via distance learning and hence does not overfit those limited concepts and intended usage patterns in MagTag5K, which leads to the better generalization on MSD. In general, we argue that our Siamese architecture should be applied in generating the *CE* representations if it can be trained on a highly informative data set, e.g., MSD in the music domain, and the computational efficiency issues arising from our raw representations can be addressed properly as discussed next.

Apparently, our proposed approach relies on the Siamese architecture of deep neural networks to learn complex contextualized semantics from descriptive terms. In general, training a deep neural network involves the non-convex optimization and tedious hyper-parameter tuning. Our proposed approach is inevitably subject to this limitation. Furthermore, we employ the $tfidf$ representation to characterize a term (used as a part of input to the deep neural network) and the BoW to represent the coherent terms in a training document (as "target" to learn the prediction). Both representations have the same number of features, equal to the size of the term vocabulary in a training data set. For a large term vocabulary in a data set, e.g., Million Song Dataset (MSD), our approach suffers from a heavy computational burden, which prohibits us from training our Siamese architecture on a data set like MSD with our current computational facility. In general, a parsimonious representation of a large word vocabulary is demanded by various natural language processing tasks and has been studied previously. The potential solutions include applying a dimensionality reduction technique, e.g., PCA or compressed censing (Hsu et al., 2009), to the representation and transforming the high-dimensional binary BoW representation into a low-dimensional continuous yet compressed representation (Hsu et al., 2009). While the potential solutions still need to be investigated, we anticipate that such techniques would effectively reduce the computational burden in our approach.

For our proposed Siamese architecture, there are several salient characteristics that distinguish our architecture from most of existing Siamese architectures. First of all, most of existing Siamese architectures are developed to learn a distance metric only in the representation space (Bordes et al., 2011; Bromley et al., 1993; Chopra, Hadsell & LeCun, 2005). Unlike those architectures that learn a single task, ours not only learns a distance metric in the CE space but also simultaneously establishes a predictor that infers the coherent terms from an instance consisting of the raw representations of a focused term and its local context. Next, the existing Siamese architectures are generally trained via supervised learning. In contrast, ours is trained in two stages via unsupervised learning. Finally, ours is also different from those regularized Siamese architectures (Chen & Salman, 2011; Salakhutdinov & Hinton, 2007). Such regularized variants employ an auto-encoder as its subnetwork in order to minimize information loss so as to achieve better generalization, while ours learns two relevant yet different tasks, prediction and distance metric learning, simultaneously. In addition, those regularized Siamese architectures are still trained via supervised learning. Apart from those salient characteristics, we believe that our proposed Siamese architecture and learning algorithms can be easily extended to other types of contextualized semantic learning from descriptive terms by means of alternative context information instead of our used local context, i.e., co-occurring terms in an annotation document, defined in this paper.

In this paper, we have formulated a contextualized semantic learning task from collections of textual descriptive terms independent of any specific MMIR application tasks. We believe that a solution to this problem would facilitate bridging the semantic gap between media content and relevant high-level concepts. Our work presented in this paper is different from the previous studies phrased with term "contextual" in this area. For example, the "contextual object recognition" (Rasiwasia & Vasconcelos, 2012) actually refers to a method of exploiting the relatedness of object labels achieved from training an object recognizer (a multi-class classifier) to improve the performance by using such information as context to train another classifier. That "context" achieved directly from media content for a specific application task is by no means relevant to our contextualized semantic learning task and the proposed solution. In addition, the "contextual tag inference" (Mandel et al., 2011) is an approach that exploits descriptive terms in order to produce a smoothed representation for documents with CRBM. The smoothed representation is a document-level summary of the document-term relatedness to improve the term-based auto-annotation performance via providing smoothed target labels instead of binary ones. This smoothed representation acts as a novel document-level representation but does not capture the term-to-term relatedness explicitly. In other words, this method does not provide an explicit continuous embedding representation for a term. Technically, this approach is subject to limitation in generalization as the learnt representation is merely applicable to the training documents due to the context characterized by the document ID. Although this limitation might be overcome by using some alternative contextual information, this method is still not a legitimate solution to our formulated problem due to the lack of continuous embedding representation. In nature, the work closest to ours is the "Association Rules" (Yang et al. 2010) that lead to contextualized semantic representations on a conceptual level. Unlike the local context used in our approach, different rules may apply different contextual information to a term for concept modeling, which could result in inconsistency due to the intricate contextual information. Furthermore, the mined rules can provide binary concept-to-concept relatedness only, which confines itself to a limited range of applications.

In general, the solution presented in this paper leads to multiple continuous *CE* representations for a descriptive term depending on the local context. In essence, one *CE* representation of a term tends to accurately model a concept intended by annotators. Moreover, the *CE* representation space and the learnt semantic distance metric allow similar concepts to associate with each other and make different concepts readily distinguished. As a result, our *CE* representation scheme significantly distinguishes from other semantic representations learnt from descriptive terms. Without taking contextual information into account, the learnt semantic representation reflects only the global term-to-term relatedness and hence each term has a unique representation (Deerwester et al., 1990; Markines et al., 2009). In the existing work in addressing contextualized semantics, most of those methods, e.g., smoothing (Mandel et al., 2011) and probabilistic topic models, LDA (Blei, Ng & Jordan, 2003) and PLSA (Hofmann, 1999), offer only a document-level representation but do not address the contextualized term-to-term relatedness issue directly. It is well known that several documents may together specify a single concept, while one document may convey multiple concepts. Therefore, a concept-level representation is always required even though an application works on the level of documents relatedness. Here, we emphasize that for a document of *m* terms, we employ all *m* concept-level representations arising from this document collectively to form a document-level representation and the learnt sematic distance metric by our Siamese architecture can easily adapt to measuring the document-to-document relatedness as demonstrated in the extended semantic priming.

Finally, it is worth mentioning that there are alternative methods for solving our formulated problem. Most of those methods fall into the ontology area and rely on human expertise such as "tag ontologies" (Kim et al., 2008) and "property lists" semantics (Sun et al., 2013). On the one hand, experts' specialist knowledge regarding terms and their inter-relatedness is harvested so that a system can use such semantics to perform human-like tasks. On the other hand, such acquisition of semantics has to be handcrafted and time-consuming, which is laborious and hence incurs a huge cost. Moreover, the acquired semantics may result in experts' bias, and the subjective opinion differences may even cause conflicting semantics. In contrast, our approach presented in this paper lends clear evidence to favor learning semantics from descriptive terms as our approach is much less costly and can automatically capture concepts underlying terms in context by following trends of the crowd in meaning and intentions.

## 8 Conclusion

We have presented an approach to acquiring contextualized semantics from co-occurring descriptive terms. In our approach, we have formulated the problem as learning a contextualized term-based semantic representation via *concept embedding* in the representation space. As a result, we have proposed a solution by developing a novel Siamese architecture of deep neural networks and a two-stage learning algorithm. We have also addressed the OOV issues in our solution. By means of visualization, we have demonstrated that our approach can capture domain-specific and transferable contextualized semantics conveyed in co-occurring terms. Moreover, we have applied our approach to semantic priming, a benchmark information retrieval task. We have conducted a thorough evaluation via a comparative study with different settings. Experimental results suggest that our approach outperforms a number of state-of-the-art approaches and the effectiveness of our proposed OOV methods in this benchmark task.

While our proposed Siamese architecture and learning algorithms provides a solution to the formulated problem, there are still several issues to be tackled including the computational efficiency in using a training corpus of a large term vocabulary; exploring alternative contextual information sources and modeling techniques; and extension of our Siamese architecture to learning other types of contextualized semantics other than the local context defined in this paper. In our ongoing work, we shall be dealing with the aforementioned issues and applying the learnt contextualized semantic representations to a number of real MMIR applications, e.g., auto-annotation of multimedia content, term/media content recommendation and query expansion, multimedia retrieval with textual queries as well as zero-shot learning in various multimedia classification tasks. We anticipate that the formulated problem and our solution presented in this paper would pave a new way towards bridging the semantic gap.

**Appendix A. Learning Algorithm Details**

In this appendix, we derive the learning algorithms used to train our proposed Siamese architecture. To minimize the loss functions defined for the prediction and the distance metric learning described in Section 4 of the main text, we apply the stochastic back propagation (SBP) algorithm for parameter estimation. To establish a deep subnetwork for the prediction learning, the pre-training is carried out in a greedy layer-wise fashion with each layer's weights obtained via training a sparse autoencoder with a Quasi-Newton method as described in Section A.1. In Section A.2 and A.3, we present the derivation of gradients of loss functions with respect to relevant parameters used to train a subnetwork for the prediction and to train the Siamese architecture for the distance metric learning, respectively. Finally, we summarize the SBP algorithm that can be used in training one subnetwork for the prediction and the Siamese architecture for the distance learning.

### A.1. Sparse Auto-encoder Learning

A sparse autoencoder can be used to initialize weights of a deep neural network by reconstructing the input via a single hidden and preferably sparse layer (Ranzato, Boureau & LeCun, 2007). In our experiments, the sparse autoencoder was trained using batch training using the Limited-memory Broyden–Fletcher–Goldfarb–Shanno (L-BFGS) method, a variant of Quasi-Newton method in the popular implementation of `minFunc` (Schmidt, 2005).

Let $x$ be an input vector. The hidden layer's activations are

$$z_1(x) = f(W_1 x + b_1),$$

and the corresponding output layer's activation are

$$\tilde{x}(x) = f(W_2 z_1(x) + b_2),$$

where $W_1$, $b_1$, $W_2$ and $b_2$ are the encoding weights, encoding biases, decoding weights and decoding biases, respectively. Note that $f(x) = \frac{e^x - e^{-x}}{e^x + e^{-x}}$ is the hyperbolic tangent function used in our experiments.

Encouraging sparsity is carried out via a regularizer to the cost which consists of penalizing the magnitude of the hidden layer's output regardless of the sign:

$$\mathcal{R} = \sum_{q=1}^{Q} \sqrt{(z_1(x)[q])^2 + \epsilon},$$

where $Q$ is the number of units in the hidden layer.

The objective of the training is minimizing the following loss averaged over all the examples:

$$\mathcal{L}_A(X; \Theta) = \sum_{k=1}^{K} \|\tilde{x}(x_k) - x_k\|_2^2 + \alpha \sum_{k=1}^{K} \left( \sum_{q=1}^{Q} \sqrt{(z_1(x)[q])^2 + \epsilon} \right).$$

Hence, we achieve $\frac{\partial \mathcal{L}_A(X;\Theta)}{\partial \tilde{x}(x)} = 2\|\tilde{x}(x) - x\|_1$ and $\frac{\partial \mathcal{R}}{\partial z_1(x)} = \frac{z_1(x)}{\sqrt{(z_1(x))^2 + \epsilon}}$.

Let $\nabla f(x)$ be the gradient of the hyperbolic function given input $x$. We have

$$\nabla f(x) = \frac{\partial f(x)}{\partial x} = \nabla \tanh(x) = 1 - (\tanh(x))^2$$

Given a training data set $X$ of $K$ examples, we applying the chain rule in order to obtain the derivatives with respect to a specific parameter as follows:

$$\frac{\partial \mathcal{L}_A(X;\Theta)}{\partial W_2} = \sum_{k=1}^{K} \frac{\partial \mathcal{L}_A(X;\Theta)}{\partial \tilde{x}(x_k)} \cdot \frac{\partial \tilde{x}(x_k)}{\partial W_2} = 2 \sum_{k=1}^{K} \left( (\tilde{x}(x_k) - x_k) \cdot \nabla f(W_2 z_1(x_k) + b_2) \cdot z_1(x_k) \right)$$

$$\frac{\partial \mathcal{L}_A(X;\Theta)}{\partial b_2} = \sum_{k=1}^{K} \frac{\partial \mathcal{L}_A(X;\Theta)}{\partial \tilde{x}(x_k)} \cdot \frac{\partial \tilde{x}(x_k)}{\partial b_2} = 2 \sum_{k=1}^{K} \left( (\tilde{x}(x_k) - x_k) \cdot \nabla f(W_2 z_1(x_k) + b_2) \right)$$

$$\frac{\partial \mathcal{L}_A(X;\Theta)}{\partial W_1} = \sum_{k=1}^{K} \frac{\partial \mathcal{L}_A(X;\Theta)}{\partial \tilde{x}(x_k)} \cdot \frac{\partial \tilde{x}(x_k)}{\partial z_1(x_k)} \cdot \frac{\partial z_1(x_k)}{\partial W_1} + \alpha \frac{\partial \mathcal{R}}{\partial z_1(x_k)} \cdot \frac{\partial z_1(x_k)}{\partial W_1}$$

$$= 2 \sum_{k=1}^{K} \left( (\tilde{x}(x_k) - x_k) \cdot \nabla f(W_2 z_1(x_k) + b_2) \cdot W_2 \cdot \nabla f(W_1 x_k + b_1) \cdot x_k \right)$$

$$+ \alpha \sum_{k=1}^{K} \left( \frac{z_1(x_k)}{\sqrt{(z_1(x_k))^2 + \epsilon}} \cdot \nabla f(W_1 x_k + b_1) \cdot x_k \right)$$

$$\frac{\partial \mathcal{L}_A(X;\Theta)}{\partial b_1} = \sum_{k=1}^{K} \frac{\partial \mathcal{L}_A(X;\Theta)}{\partial \tilde{x}(x_k)} \cdot \frac{\partial \tilde{x}(x_k)}{\partial z_1(x_k)} \cdot \frac{\partial z_1(x_k)}{\partial b_1} + \alpha \frac{\partial \mathcal{R}}{\partial z_1(x_k)} \cdot \frac{\partial z_1(x_k)}{\partial b_1}$$

$$= 2 \sum_{k=1}^{K} \left( (\tilde{x}(x_k) - x_k) \cdot \nabla f(W_2 z_1(x_k) + b_2) \cdot W_2 \cdot \nabla f(W_1 x_k + b_1) \right)$$

$$+ \alpha \sum_{k=1}^{K} \left( \frac{z_1(x_k)}{\sqrt{(z_1(x_k))^2 + \epsilon}} \cdot \nabla f(W_1 x_k + b_1) \right) \quad \text{(A.1)}$$

The sparse auto-encoder is employed to initialize a subnetwork recursively where each layer is trained based on the output of its previous layer all the way until a pre-specified number of layers are achieved.

## A.2. Subnetwork Learning for Prediction

As defined in Equation 5 of the main text, the prediction loss is

$$\mathcal{L}_P(X;\Theta) = -\frac{1}{2K|\Gamma|}\sum_{k=1}^{K}\mathcal{L}_P(x_k;\Theta),$$

where $\mathcal{L}_P(x_k;\Theta) = \sum_{i=1}^{|\Gamma|}(\kappa_k(1+y_k[i])\log(1+\hat{y}_k[i]) + (1-\kappa_k)(1-y_k[i])\log(1-\hat{y}_k[i]))$.

Here, $\hat{y}_k[i]$ and $y_k[i]$ represent the prediction and the true label related to term $i$ in example $k$, respectively. By applying the chain rule, we have

$$\frac{\partial \mathcal{L}_P(X;\Theta)}{\partial x} = \frac{1}{K}\sum_{k=1}^{K}\left(\frac{\partial \mathcal{L}_P(x_k;\Theta)}{\partial \hat{y}_k} \cdot \frac{\partial \hat{y}_k}{x}\right),$$

where $\hat{y}_k$ is the output vector of prediction, a collective notation of all $\hat{y}_k[i]$, also . operator is the element wise multiplication. We have

$$\frac{\partial \mathcal{L}_P(x_k;\Theta)}{\partial \hat{y}_k} = \frac{-1}{2|\Gamma|}\sum_{i=1}^{|\Gamma|}\left(\kappa_k \frac{1+y_k[i]}{1+\hat{y}_k[i]} \cdot \frac{\partial \hat{y}_k[i]}{\partial \hat{y}_k} - (1-\kappa_k)\frac{1-y_k[i]}{1-\hat{y}_k[i]} \cdot \frac{\partial \hat{y}_k[i]}{\partial \hat{y}_k}\right)$$

$$= \frac{-1}{2|\Gamma|}\left(\kappa_k \frac{1+y_k}{1+\hat{y}_k} - (1-\kappa_k)\frac{1-y_k}{1-\hat{y}_k}\right),$$

where $\frac{1+y}{1+\hat{y}}$ and $\frac{1-y}{1-\hat{y}}$ are the collective notations of $\frac{1+y_k[i]}{1+\hat{y}_k[i]}$ and $\frac{1-y_k[i]}{1-\hat{y}_k[i]}$ with the element-wise division.

Let $\frac{\partial \mathcal{L}_P(X;\Theta)}{\partial \hat{Y}}$ be the matrix formed by stacking all training $\frac{\partial \mathcal{L}_P(x_k;\Theta)}{\partial \hat{y}_k}$ and $z_{H-1}(X)$ be the matrix formed by stacking all $z_{H-1}(x_k)$. Back propagation starts at the top layer with the partial of the cost on the last layer's parameters (weights and biases), $\frac{\partial \mathcal{L}_P(X;\Theta)}{\partial W_H}$ and $\frac{\partial \mathcal{L}_P(X;\Theta)}{\partial b_H}$, given by

$$\frac{\partial \mathcal{L}_P(X;\Theta)}{\partial W_H} = \frac{1}{K}\sum_{k=1}^{K}\left(\frac{\partial \mathcal{L}_P(x_k;\Theta)}{\partial \hat{y}_k} \cdot \nabla f(W_H z_{H-1}(x_k) + b_H) \cdot \frac{\partial (W_H \times z_{H-1}(x_k) + b_H)}{W_H}\right)$$

$$= \frac{1}{K}\left(\frac{\partial \mathcal{L}_P(X;\Theta)}{\partial \hat{Y}} \cdot \nabla f(W_H z_{H-1}(X) + b_H)\right) * \left(z_{H-1}(X)\right)^T,$$

$$\frac{\partial \mathcal{L}_P(X;\Theta)}{\partial b_H} = \frac{1}{K}\sum_{k=1}^{K}\left(\frac{\partial \mathcal{L}_P(x_k;\Theta)}{\partial \hat{y}_k} \cdot \nabla f(W_H z_{H-1}(x_k) + b_H)\right)$$

$$= \frac{1}{K}\left(\frac{\partial \mathcal{L}_P(X;\Theta)}{\partial \hat{Y}} \cdot \nabla f(W_H z_{H-1}(X) + b_H)\right), \tag{A.2}$$

where $*$ is the matrix multiplication.

Derivatives with respect to all the parameters, $W_h$ and $b_h$, of hidden layer $h = (H-1,\dots,1)$ are obtained by the successive use of the chain rule for error back-propagation:

$$\frac{\partial \mathcal{L}_P(X;\Theta)}{\partial W_h} = \underbrace{\frac{\partial \mathcal{L}_P(X;\Theta)}{\partial (W_h z_{h-1}(X)+b_h)}}_{\zeta_h} \cdot \frac{\partial (W_h z_{h-1}(X)+b_h)}{\partial W_h},$$

$$\zeta_h = \zeta_{h+1} \cdot \frac{\partial (W_{h+1} z_h(X)+b_{h+1})}{\partial z_h(X)} \cdot \frac{\partial z_h(X)}{\partial (W_h z_{h-1}(X)+b_h)} = W_{h+1} * \left(\zeta_{h+1} \cdot \nabla f(W_h z_{h-1}(X)+b_h)\right),$$

$$\frac{\partial (W_h z_{h-1}(X)+b_h)}{\partial W_h} = z_{h-1}(X),$$

$$\frac{\partial \mathcal{L}_P(X;\Theta)}{\partial b_h} = \frac{\partial \mathcal{L}_P(X;\Theta)}{\partial (W_h z_{h-1}(X)+b_h)} \cdot \frac{\partial (W_h z_{h-1}(X)+b_h)}{\partial b_h} = \zeta_h. \tag{A.3}$$

### A.3. Siamese architecture Learning

As defined in Equation 6 of the main text, the Siamese loss is

$$\mathcal{L}_S(X^{(1)}, X^{(2)}; \Theta) = \frac{1}{K}\sum_{k=1}^{K}\left(I_1(\mathbb{E} - \beta(1 - \mathbb{D}))^2 + I_2\rho(\mathbb{E} - \beta(1 - \mathbb{D}))^2 + I_3(\mathbb{E} - \beta)^2\mathbb{D}\right).$$

Here $\mathbb{E} = E(x_k^{(1)}, x_k^{(2)})$ is the Euclidian distance between the embedding vectors of pairs of input examples and $\mathbb{D} = e^{-\frac{\lambda}{2}KL(x_k^{(1)}, x_k^{(2)})}$ the target distance is based on contexts similarity following $KL(x^{(1)}, x^{(2)}) = \sum_{c=1}^{|\Phi|}\left((l^{(1)}[c] - l^{(2)}[c])\log\left(\frac{l^{(1)}[c]}{l^{(2)}[c]}\right)\right)$ where $|\Phi|$ features in the context representation and $l^{(i)}[c]$ represents the $c^{th}$ feature value in the context input provided for subnetwork number $i$. Note that $\beta(1 - \mathbb{D})$ is a constant irrespective of the weights and biases. Moreover, this loss is also unaffected by any weights connected between the CE (i.e., hidden layer $H - 1$) and the prediction layers. Thus, $\frac{\partial \mathcal{L}_S(X^{(1)}, X^{(2)}; \Theta)}{\partial W_H} = 0$ and $\frac{\partial \mathcal{L}_S(X^{(1)}, X^{(2)}; \Theta)}{\partial b_H} = 0$.

As two subnetworks always need to be kept identical, all the parameters in each subnetwork are updated by using the averaging derivatives obtained based two subnetworks after each back propagating iteration. As there is no interaction between the two subnetworks apart from the CE layers, we can write

$$\frac{\partial \mathcal{L}_S(x_k^{(1)}, x_k^{(2)}; \Theta)}{\partial \Theta} = \frac{1}{2}\left(\frac{\partial \mathcal{L}_S(x_k^{(1)}, x_k^{(2)}; \Theta)}{\partial CE(x_k^{(1)})} \cdot \frac{\partial CE(x_k^{(1)})}{\partial \Theta} + \frac{\partial \mathcal{L}_S(x_k^{(1)}, x_k^{(2)}; \Theta)}{\partial CE(x_k^{(2)})} \cdot \frac{\partial CE(x_k^{(2)})}{\partial \Theta}\right).$$

As the loss is symmetric in terms of the embedding vectors, the derivatives have a uniform form for subnetworks $i = 1, 2$:

$$\frac{\partial \mathcal{L}_S(X^{(1)}, X^{(2)}; \Theta)}{\partial CE(X^{(i)})} = \frac{1}{K}\sum_{k=1}^{K}\frac{\partial \mathcal{L}_S(x_k^{(1)}, x_k^{(2)}; \Theta)}{\partial CE(x_k^{(i)})}$$

$$\frac{\partial \mathcal{L}_S(x_k^{(1)}, x_k^{(2)}; \Theta)}{\partial CE(x_k^{(i)})} = 2\left(I_1(\mathbb{E} - \beta(1 - \mathbb{D})) + \rho I_2(\mathbb{E} - \beta(1 - \mathbb{D})) + I_3((\mathbb{E} - \beta)\mathbb{D})\right) \cdot \frac{\partial E(x_k^{(1)}, x_k^{(2)})}{\partial CE(x_k^{(i)})}$$

Focusing on $\frac{\partial E(x_k^{(1)}, x_k^{(2)})}{\partial CE(x_k^{(i)})}$, we have

$$\frac{\partial E(x_k^{(1)}, x_k^{(2)})}{\partial CE(x_k^{(i)})} = \frac{\|CE(x_k^{(1)}) - CE(x_k^{(2)})\|_1}{E(x_k^{(1)}, x_k^{(2)})}; \quad E(x_k^{(1)}, x_k^{(2)}) = \|CE(x_k^{(1)}) - CE(x_k^{(2)})\|_2.$$

Effectively, we can now estimate the partial derivatives for the embedding layer's weights and biases regarding subnetworks $i = 1, 2$:

$$\frac{\partial \mathcal{L}_S(X^{(1)}, X^{(2)}; \Theta)}{\partial W_{H-1}^{(i)}} = \frac{1}{K}\sum_{k=1}^{K}\frac{\partial \mathcal{L}_S(x_k^{(1)}, x_k^{(2)}; \Theta)}{\partial CE(x_k^{(i)})} \cdot \frac{\partial CE(x_k^{(i)})}{\partial W_{H-1}^{(i)}}$$

$$= \frac{1}{K}\sum_{k=1}^{K}\left(\frac{\partial \mathcal{L}_S(x_k^{(1)}, x_k^{(2)}; \Theta)}{\partial CE(x_k^{(i)})} \cdot \nabla f(W_{H-1}z_{H-2}(x_k^{(i)}) + b_{H-1})\right) \cdot \frac{\partial(W_{H-1}z_{H-2}(x_k^{(i)}) + b_{H-1})}{\partial W_{H-1}}$$

$$= \frac{1}{K}\left(\frac{\partial \mathcal{L}_S(x_k^{(1)}, x_k^{(2)}; \Theta)}{\partial CE(x_k^{(i)})} \cdot \nabla f(W_{H-1}z_{H-2}(x_k^{(i)}) + b_{H-1})\right) * \left(z_{H-2}(x_k^{(i)})\right)^T,$$

$$\frac{\partial \mathcal{L}_S(X^{(1)}, X^{(2)}; \Theta)}{\partial b_{H-1}^{(i)}} = \frac{1}{K}\sum_{k=1}^{K}\left(\frac{\partial \mathcal{L}_S(x_k^{(1)}, x_k^{(2)}; \Theta)}{\partial CE(x_k^{(i)})} \cdot \nabla f(W_{H-1}z_{H-2}(x_k^{(i)}) + b_{H-1})\right). \quad (A.4)$$

The rest of the derivatives are obtained by back propagating the cost in the same fashion as presented in Equations A.3.

## A.4. Stochastic Gradient Descent Procedure

Here, we present a generic stochastic gradient descent (SGD) procedure applicable to training a subnetwork for the prediction and the Siamese architecture with the derivatives in Equations A.2-A.4. Given a training data set $(X, Y)$ where $X$ is the set of input instances consisting of $tfidf$ and context features and $Y$ is the set of corresponding documents represented in the BoW, the SGD procedure is summarized as follows:

| **Algorithm: Stochastic Gradient Descent for Loss $\mathcal{L}(f(X), Y; \Theta)$** |
|---|
| **Input:** Initial parameters $\Theta_0$, initial learning rate $\eta_0$ and a stopping threshold $e_T$. |
| **Output:** Optimal parameters $\Theta_T$ |
| 1:     $t \leftarrow 1$ |
| 2:     **Repeat** |
| 3:          $\widehat{Y}_t \leftarrow f(X; \Theta_t)$ |
| 4:          $e_t \leftarrow \mathcal{L}(\widehat{Y}_{(t)}, Y; \Theta_t)$ |
| 5:          $\nabla \mathcal{L}(t) \leftarrow$ gradient of $\mathcal{L}(\widehat{Y}_t, Y; \Theta_t)$ |
| 6:          $\Theta_{t+1} \leftarrow \Theta_t - \eta_t \cdot \nabla \mathcal{L}(t)$ |
| 7:          $\eta_{t+1} \leftarrow \begin{cases} 0.95 * \eta_t & \text{if } (t \bmod 200) = 0 \\ \eta_t & \text{otherwise} \end{cases}$ |
| 8:          $t \leftarrow t + 1$ |
| 9:     **Until** $\frac{e_{t-1}}{e_t} < e_T$ |
| 10:    $\Theta_T \leftarrow \Theta_t$ |

It is worth clarifying that the stopping condition in the SGD is generic and applicable to any applications. However, we used a specific stopping condition in our experiments as described in Section 5.2.